\documentclass[%
aip,
 jmp,%
 amsmath,amssymb,
preprint,%
]{revtex4-1}

\usepackage{graphicx}

\usepackage{dcolumn}%
\usepackage{bm}%

\def\ltsima{$\; \buildrel < \over \sim \;$}
\def\gtsima{$\; \buildrel > \over \sim \;$}
\def\simlt{\lower.5ex\hbox{\ltsima}}
\def\simgt{\lower.5ex\hbox{\gtsima}}

\usepackage{color}

\begin{document}
\preprint{}

\title{Magnetic reconnection and plasmoid formation driven by the two-dimensional resistive kink instability  
in a current hole configuration}%
\author{ Hubert BATY}
\email{hubert.baty@unistra.fr}

\affiliation{Observatoire Astronomique de Strasbourg, Universit\'e de Strasbourg, CNRS, UMR 7550, 11 rue de l'Universit\'e, F-67000 Strasbourg, France}
\date{\today}%

\begin{abstract}
We investigate the nonlinear evolution of the $m = 1$ resistive internal kink mode in a two-dimensional (2D) configuration containing a central
region of negative current density, also known as the ``current hole" setup. The finite-element code FINMHD is used to solve a reduced
set of incompressible Magnetohydrodynamic (MHD) equations with a current-vorticity formulation. First, the kink instability linearly develops 
in agreement with the general theory of resistive internal kink mode, and it subsequently leads to the formation of a current sheet.
At relatively low Lundquist number $S$, a magnetic reconnection process proceeds with a rate predicted by the
Sweet-Parker regime. Conversely, when $S$ exceeds a critical value that is $S_c \simeq 10^4$, the current sheet is disrupted by the formation of
plasmoids on a slightly sub-Alfv\'enic time scale. In the latter case, a stochastic reconnection regime exhibiting Petschek-type features
enriched by plasmoids is reached. A relatively fast normalized reconnection rate value of order $0.02$ is also measured. Finally, we compare our
results with those obtained in similar 2D previous studies using ideal MHD instabilities to initiate the process, and discuss their relevance 
for the general theory of plasmoid chains formation and associated fast reconnection regime.

 \end{abstract}


\maketitle

\clearpage

\section{Introduction}

It is actually well admitted that the presence of small magnetic islands in strongly magnetized plasmas, usually called plasmoids, is of
fundamental importance in order to achieve a fast magnetic reconnection process. Magnetic reconnection is thought to be the
underlying mechanism that powers explosive events in many plasmas (e.g. solar/stellar flares in astrophysical environments or disruptions in
plasma-fusion laboratory experiments). In this way, a fraction of the free magnetic energy stored in the system can be released and converted into other
forms of energy, like kinetic bulk flow, particles acceleration, and heat. 
In magnetohydrodynamic framework approximation, the heart of the mechanism is provided by 
spatially localized layers of intense electrical current accumulation leading to the reconnection of field lines due to a
small non zero resistivity in these highly conducting plasmas.
Thus, such current sheets are thin resistive layers controlling the change
of magnetic field connectivity between two regions of oppositely directed magnetic fields in two dimensions (or having a reversal component in three dimensions). \cite {priest00,bisk09}

These current sheets may spontaneously form as the result of the development of some magnetohydrodynamic (MHD) instabilities. 
This is for example the case in two-dimensional (2D) ideally unstable setups like, the coalescence between attracting flux bundles,\cite {hua10}
and the tilt instability in repelling current channels. \cite {richard90} In the two latter
configurations, the formation process of the current sheet is consequently achieved on a purely Alfv\'enic time scale (i.e. independent of the diffusion parameters).
The route leading to fast reconnection is only attained during a second stage if the Lundquist number is higher $S$ than a critical value (hereafter $S_c$),\cite {lou07} otherwise
a Sweet-Parker (SP) reconnection subsequently ensues with a normalized reconnection rate scaling as $S^{-1/2}$. Note that SP regime is analogous to
Kadomstev model for magnetic reconnection associated to internal kink mode in tokamaks. \cite {par57, kado75} This important Lundquist number $S$
is generally defined as, $S = L V_A/\eta$, where $L$ is the half-length of the current sheet, $V_A$ is the Alfv\'en velocity based on the magnetic field amplitude
of the reversal component (upstream of the current layer), and $\eta$ being the magnetic diffusivity (i.e. resistivity).
The value of $S_c$ that is often quoted in the literature is $S_c \sim 10^4$ (see for example Ref. 8).
However, there is no precise value, as when it is obtained via numerical experiments, it depends on numerical noise. It seems to also 
depend on different plasma parameters like 
the plasma-$\beta$,  \cite {ni12} and the magnetic Prandtl number $P_r$.  \cite  {comi16} The effect of the choice of the initial
configuration is also not clearly established.
For example,values of  $S_c \simeq 3 \times 10^4$ and $ S_c \simeq 5 \times 10^3$ have been obtained for coalescence setup (with zero viscosity) ant tilt instability setup
(with $P_r = 1$) respectively.  \cite {hua10, hua17, baty19, bat20a, bat20b} 
A value of $S_c \simeq 8000$ has been also reported using coalescence instability setup in a recent numerical study.  \cite {rip19} 
If $S \simgt S_c$, the formation of plasmoid chains is triggered, as many small secondary resistive-type tearing instabilities 
develop and disrupt the current layer in which they are born. \cite {lou07} Thus, this constitutes the second stage of the scenario 
occurring for plasmas under consideration in the present study, as the relevant Lundquist number can attain large/huge values
(e.g.  $S \sim 10^{8}$ and $S \sim 10^{12}$ for tokamaks and solar corona respectively). \cite {priest00,bisk09}

The growth of the plasmoids observed in MHD simulations using coalescence/tilt setups is surprisingly fast, as the growth rate $\gamma_p$ can
easily attain super-Alfv\'enic values for $S >> S_c$ , e.g. $\gamma_p \tau_A  \simeq 10-20$ where $\tau_A$ is the Alfv\'en time defined as $\tau_A = L/V_A$.  \cite {hua17, bat20a, bat20b}
These values seem to agree with the scenario proposed by Comisso et al., \cite {com16, com17} where an analytical model based on a modal stability calculation
of such forming current sheet is developed. In the latter model, the linear growth rate of the dominant mode appears as a not simple $S$-power law, thus
offering a possible issue to the paradoxal result of infinite linear growth rate obtained in the infinitely high-$S$ limit
for static pre-formed SP current sheets (i.e. having a large aspect ratio $L/a \simeq S^{1/2}$), where $\gamma_p \tau_A \propto S^{1/4}$.
Note that, another issue has been proposed by considering unstable
current layers having a critical aspect ratio $L/a \simeq S^{1/3}$, that is substantially smaller than
SP value in the high S limit. \cite {puc14, puc18} Indeed, in this way, it is shown that the linear growth rate becomes
close to Alfv\'enic and independent of the Lundquist number, as for example $\gamma_p \tau_A  \simeq 0.62$ for Harris-type configuration with zero viscosity.
{\cite{landi15} The latter instability was thus called the ``ideal" tearing mode in the latter studies.
As these two theoretical models differ in the assumptions concerning the formation mechanism  of the current layer,
it would be of interest to consider other type of MHD instabilities for the initial step, motivating the particular choice of a resistive
mode in the present study. Resistive-type instabilities generally lead to a linear growth rate scaling as a negative power of the Lundquist
number, \cite {bisk09, white91} e.g. $S^{-1/3}-S^{-3/5}$. 

The question of the understanding of the third stage, which is the subsequent regime of fast magnetic reconnection allowed by 
the presence of plasmoid chains, is also actually under debate. The plasmoids are constantly forming, moving, eventually coalescing,
and finally being ejected through the outflow boundaries. At a given time, the system contains aligned layer structures of plasmoids
of different sizes, and can be regarded as a statistical steady state with a time-averaged normalized reconnection rate value of order $0.01-0.02$ that is quasi
-independent of the dissipation parameters \cite {uzd10, lou12}.
Fractal models (for the hierarchical structure of the plasmoid chains observed in simulations) based on heuristic arguments have been proposed
to explain this fast rate independent of the Lundquist number \cite {uzd10, ji11}. Indeed, a simple picture is used, where
as the plasmoid instability proceeds, the plasmoids grow in size and the currents sheets between primary plasmoids are again unstable. These
secondary current sheets are thinner than the primary ones and give rise to secondary plasmoids, which eventually lead to tertiary current sheets,
and so on, as originally envisioned by Shibata and Tanuma (2001). \cite {shi01} This process of multiple stages of cascading ends up when the thinnest current
structures between plasmoids are short enough, i.e. corresponding to marginally stable Sweet-Parker layers. However, a close inspection of the
MHD simulations indicates that some of the fragmented current sheets do not have enough time to form SP layers. \cite {hua10}
and a non negligible coalescence effect between adjacent plasmoids leads to the formation of so called monster plasmoids. \cite {lou12}
Moreover, in a recent numerical study using the tilt setup at high enough $S$ values (typically for $S \sim 100S_c$), a dynamical Petschek-type reconnection
is shown to be achieved  with pairs of slow-mode shocks emanating from a small central region containing a few plasmoids. \cite {bat20c} The role played by
merging events of plasmoids giving rise to monster plasmoids with shocks bounding the outflow regions is fundamental to this respect. \cite {bat20c}

The goal of the present work is precisely to address the role of the initial formation mechanism of the current sheet on the subsequent
plasmoid chains formation and also on the associated fast magnetic reconnection process. In order to achieve this aim, the results
obtained for a resistive-type 2D MHD instability (e.g. an internal kink mode) will be compared in particular to results from
our previous studies using the ideal tilt mode.  \cite {bat20a, bat20b, bat20c}
The configuration considered in this study is known in the literature as the ``current hole" setup and has been
used in the context of the MHD stability of advanced tokamak scenarios with reversed central current. \cite {huy01,cza08}

The outline of the paper is as follows. In Sect. II, we present the MHD code and the initial setup 
for the 2D resistive kink instability. Section III is devoted to the presentation of the results. Finally, we discuss our results
and conclude in Sect. IV.

\section{The MHD code and initial setup}

\subsection{FINMHD equations}
A set of reduced MHD equations has been employed corresponding to a 2D incompressible model. However,
instead of taking the usual formulation with vorticity and magnetic flux functions for the main variables, another choice
using current-vorticity ($J-\omega$) variables is preferred because of its more symmetric formulation, facilitating the numerical
matrix calculus. The latter choice also cures numerical difficulty due to the numerical
treatment of a third order spatial derivative term. \citep{baty19} To summarize, the following set of equations is,
\begin{equation}  
      \frac{\partial \omega}{\partial t} + (\bm{V}\cdot\bm{\nabla})\omega = (\bm{B}\cdot\bm{\nabla})J + \nu \bm{\nabla}^2 \omega ,
\end{equation}
\begin{equation}      
        \frac{\partial J }{\partial t} + (\bm{V}\cdot\bm{\nabla})J =  (\bm{B}\cdot\bm{\nabla})\omega + \eta \bm{\nabla}^2 (J - J_e) +  g(\phi,\psi) ,
\end{equation}
\begin{equation}                 
     \bm{\nabla}^2\phi = - \omega ,
 \end{equation}
\begin{equation}                        
     \bm{\nabla}^2\psi = - J ,  
\end{equation}
with $g(\phi,\psi)=2 \left[\frac{\partial^2 \phi}{\partial x\partial y}\left(\frac{\partial^2 \psi}{\partial x^2} - \frac{\partial^2 \psi}{\partial y^2}\right) - \frac{\partial^2 \psi}{\partial x\partial y}\left(\frac{\partial^2 \phi}{\partial x^2} - \frac{\partial^2 \phi}{\partial y^2}\right)\right]$.
As usual, we have introduced the two stream functions, $\phi (x, y)$ and $\psi (x, y)$, from the fluid velocity $\bm{V} = {\nabla} \phi \wedge \bm{e_z}$ and magnetic field $\bm{B} = {\nabla} \psi \wedge \bm{e_z}$, $\bm{e_z}$
being the unit vector perpendicular to the $xOy$ simulation plane.
$J$ and vorticity $\omega$ are the $z$ components of the current density and vorticity vectors, as $\bm{J} = \nabla \wedge \bm{B}$ and $\bm{ \omega} = \nabla \wedge \bm{V}$ respectively (with units using $\mu_0 = 1$).
Note that we consider the resistive diffusion via the $\eta \bm{\nabla}^2 J $ term ($\eta$ being the resistivity that is assumed uniform for simplicity), and also a viscous term
$\nu \bm{\nabla}^2 \omega$ in a similar way (with $\nu$ being the viscosity parameter also assumed uniform).
The above definitions results from the choice $\psi \equiv A_z$ for the magnetic flux function, where $A_z$ is the $z$ component of the potentiel vector $\bm{A}$, as $\bm{B} = \nabla \wedge \bm{A}$. $J_e (r)$ is the initial
current density profile, which is introduced in Eq. 2 via the $ - \eta \bm{\nabla}^2 J_e$ term in order to compensate the purely resistive dissipation
of the current equilibrium. Contrary to previous studies using the tilt setup where it was absent,
this extra-term is important when a resistive instability is considered.
Note that thermal pressure gradient is naturally absent from our set of equations. 
Finally, an advantage of the above formulation over a standard one using the velocity and magnetic field vectors ($\bm{V}, \bm{B}$)  as
the main variables, is the divergence-free property naturally ensured for these two vectors. 

The reduced set of MHD equations allows us to simplify and avoid having to solve the full MHD
equations. This is fully justified in situations where the magnetic field is dominated by a strong longitudinal/toroidal
component like in a tokamak device. Indeed, it is a good approximation for the 2D dynamics perpendicular to the longitudinal
direction that is essentially incompressible. \cite {bisk09} Moreover, this also facilitates the comparison of our results with previous ones
on the 2D resistive kink mode where this choice was also done. \cite {huy01,cza08}

\subsection{FINMHD numerical method}

FINMHD code is based on a finite-element method using triangles with quadratic
basis functions on an unstructured grid. A characteristic-Galerkin scheme is chosen 
in order to implement in a stable way the Lagrangian derivative $\frac{\partial  }{\partial t} + (\bm{V}\cdot\bm{\nabla}) $ appearing
in the two first equations.
Moreover, a highly adaptive (in space and time) scheme is developed in order to follow the rapid
evolution of the solution, using either a first-order time integrator (that is linearly unconditionally stable) or a second-order one
(subject to a Courant-Friedrichs-Lewy time-step restriction). Typically, a new adapted grid based on the Delaunay-Voronoi algorithm
can be computed at each time step, by searching the grid that renders an estimated error nearly uniform. 
Practically, a maximum value for the elements edge size $h_{max}$ is chosen in order to capture
the solution in regions where it behaves smoothly. $h_{max}$ thus represents a base resolution. Moreover, the grid mesh
is adapted in time to the solution by using the Hessian matrix of the current density. In this way, a non uniform mesh
can be obtained by generating smaller triangles where it is needed (e.g in current sheets). A value 
is initially specified for the maximum number of elements (i.e. triangles) $n_t$.
But the latter constraint is practically never reached at any time, as $n_t \simeq 600000$ is very high and never reached in this work.
Our adaptive mesh procedure is shown to efficiently capture the small-scale features with a few tens of triangles
across the solution (see Appendix B). Additionally, convergence studies are carried out by limiting the smallest
edge size of the triangles $h_{min}$ (see Appendix C).
The time step can vary and is adapted from an initial value chosen to accurately follow the linear phase of the instability, and from
the current status of the system evolution by taking the maximum current density amplitude variation.
The technique used in FINMHD has been tested on challenging tests, involving 
unsteady strongly anisotropic solution for the advection equation, formation of shock structures
for viscous Burgers equation, and magnetic reconnection process for our reduced set of MHD equations.
The finite-element Freefem++ software allows to do this, in a simple and  efficient way. \citep{hec12}
The reader may refer to Baty (2019) for more details on the numerical scheme.  \citep{baty19}

\subsection{The initial setup}

In this work, we consider MHD equilibria consistent with radial current density profiles of the form,
\begin{equation}
    J_e = J_1 (1 - \tilde{r}^\alpha) - J_2 (1 - \tilde{r}^2)^8,
  \end{equation}
where $\tilde{r}$ is the normalized radius, i.e. $\tilde{r} = r/a$ with $a$ being the outer boundary radius.
In this study, distance is normalized to the value of $a$ so that $a = 1$.
Such equilibria with $\alpha = 4$ were used in the context of tokamak scenarios with reversed
central current in order to study the current hole configuration. \cite {huy01,cza08}
In the present study, we have preferred
to use a slightly smoother profile with $\alpha = 2$. Moreover, we have chosen $J_1 = 20$ and $J_2 = 36$
defining thus our normalization. In this way, the maximum current density has a similar value compared
to the tilt setup one (where the maximum equilibrium current density was of order $10$), as
one can see in Fig. 1 (see also Appendix A). Note that the current hole (negative current density region) is consequently situated in a
core region $r \simlt 0.28 a$. The corresponding azimuthal magnetic field component $B_\theta (r)$, that is also plotted in Fig. 1,
exhibits a direction reversal at the specific radius, $r _s \simeq 0.43 a$. One must note that, any perpendicular magnetic
field is not contributing to the present 2D equilibrium, and should be consequently ensured via an extra thermal pressure term (not needed in
our reduced model).

In a related three-dimensional (3D) cylindrical geometry, the specific radius $r_s$ would represent the resonant $q =  \infty$
surface, where $q(r)$ is the usually defined safety factor as $q (r)  \propto B_0/B_\theta$, with $B_0$
being a constant toroidal/axial magnetic field component. \cite {huy01,cza08}
It also defines the initial radial localization of the current layer, as
it corresponds to a local maximum of the corresponding equilibrium magnetic flux $\psi_e (r)$ (obtained by
solving the second Poisson equation of our model), as one
can see in Fig. 2. Indeed, 
such non monotonic flux function completely determines the the magnetic flux to be reconnected on each side
of the specific/resonant radius $r_s$, where the radial flux derivative vanishes (i.e. $ \psi_e^ {'} (r_s) = 0$).
The instability giving rise to the reconnection process is a resonant mode and corresponds to
$q (r_s) = m/n$, with the toroidal mode number $n$ equal to $0$ because of the axisymmetry of the configuration,  $m$ being the
poloidal mode number. Figure 2 also indicates how two magnetic field lines situated at different radius (see the dashed lines)
on each side of the resonant radius are going to reconnect, the last reconnection event
occurring between $r = 0$ and $r  \simeq 0.7 a$ radii. The dominant instability
is also shown to be the $m =1$ internal kink, where $m$ is the poloidal/azimuthal mode number (see below).
In spite of some common features, the present $m = 1$ resistive kink instability must be distinguished from its ``cousin", the
$m/n = 1/1$ mode developing in 3D cylindrical tokamak approximation because of the presence of a resonant $q = 1$ surface,
and where our azimuthal magnetic field $B_\theta$ is replaced by the helical magnetic field component
$B_h = (1 - q)  B_\theta$ (see Ref. 2).  \cite {bisk09}

Note the difference in our normalization by a factor of $100$ compared to previous papers,  \cite {huy01,cza08} leads to a magnetic field
value of order unity in the radial region close to $r_s$ (see Fig. 1). In this way, the Alfv\'en time based on $B_\theta$ is thus $t_A = a / V_A \simeq 1$.
As a consequence employing a resistivity parameter of $\eta = 10^{-4}$ (in our case) is equivalent of using  $\eta = 10^{-6}$ in the previous studies,  \cite {huy01,cza08}
in order to get the same Lundquist number $S^* = a V_A /\eta = 10^4$. $S^*$ is just the inverse of the resistivity value here, and must be distinguished
from the other Lundquist number $S$ defined in introduction and discussed further in this paper.

Instead of imposing an explicit small amplitude function in order to perturb the initial setup, we have chosen to let the instability develops from
the numerical noise. Consequently, an initial zero stream function is assumed $ \phi_e (x, y) = 0$,
with zero initial vorticity $ \omega_e (x, y) = 0$.  The values of our four different variables are also
imposed to be constant in time and equal to their initial values at the boundary $r = a = 1$.

\section{MHD Simulation Results}
\label{results}

\subsection{Typical evolution}

The typical behavior of our unstable configuration is illustrated in Fig. 2, with snapshots of the current density overlaid with magnetic
field lines. The resistivity value employed for this run is  $\eta = 2 \times 10^{-3}$ with a magnetic Prandtl value $P_r = \nu / \eta =
1$. The well known off-axis displacement of the core region situated inside $r_s$ (i.e the radial step-like function typical of the $m = 1$ kink mode)
is particularly visible in snapshot (b). The direction of the displacement is not a priori determined as no specific perturbation is imposed,
and it thus can vary from case to case in the different runs described in the present study.
The ensuing magnetic reconnection process is at work with a growing in time $m = 1$ magnetic island (at the expense of the shifted core)
in the snapshots (c)-(e), which is driven by
the curved current sheet with an also (growing in time) intense positive current density. Finally, reconnection ends up when the $m = 1$ island
containing the reconnected field lines invade the whole central region. The corresponding time evolution of the
two main variables ($J$ and $\Omega$) are also plotted in Fig. 3. Indeed, we have reported the maximum amplitudes taken over the whole domain
of the current density,  $J_{max}$ and vorticity $\Omega_{max}$. One can clearly see the early development of the instability starting from the numerical noise,
due to the initial discretization, for $\Omega_{max}$ at $ t \simeq 6 t_A$. Then a quite long linear phase of vorticity growth in association with the $m = 1$ displacement
is followed by the formation of a current sheet at $t \simeq 18 t_A$, subsequently driving the reconnection until $t \simeq 22.5 t_A$ (i.e. at
the time of the peak amplitude).

The final state is obtained after a fast relaxation leading to a configuration with a monotonic magnetic flux free of current density and
magnetic field reversals (see Fig. 4). The maximum current density is consequently reaching a constant value close to the initial one (see Fig. 3).
Thus, the evolution of the $ m = 1$ resistive kink mode in our 2D current hole configuration 
is very similar in all aspects to the evolution of the internal $m/ n = 1/1$ kink mode obtained in 3D dimensional cylindrical geometry.

\subsection{Linear phase of the resistive kink} 

First, we have investigated the scaling of the linear growth rate of the kink mode with the resistivity for a fixed
magnetic Prandtl number $P_r = 1$. Different equivalent diagnostics can be used even if the velocity or kinetic energy associated
to the instability are generally taken in the literature. As the vorticity is a main variable in our model, it is thus more natural
to use it in order to evaluate the growth rate. Indeed, the slope of the vorticity evolution (semi-log plot) can be directly
used to this aim, as illustrated in Fig. 5. In this way, the normalized linear growth rate $\gamma t_A$ is computed for a large range
of resistivity values investigated in this work, and is plotted in panel (a) of Fig. 6 as a function of the resistivity for a fixed
magnetic Prandtl value $P_r = 1$.
We have found a scaling $\gamma t_A \propto  \eta^{1/3}$ in the small resistivity limit. This result is in perfect agreement with
previous ones reported for 2D resistive kink in reduced MHD framework
using the current density profile $\alpha = 4$, \cite {huy01,cza08}
and for the 3D cylindrical resistive kink.  \cite {men18}
 This is also in agreement with the scaling law expected from the theory of internal kink mode in cylindrical geometry. \cite {bisk09}
As shown in panel (b) of Fig. 6, we have also investigated the viscosity dependence of the linear growth rate by varying the magnetic
Prandtl number at a fixed resistivity value $\eta = 10^{-3}$. Again, our results fully agree with the scalings presented in the recent work
for the 3D $m/ n = 1/1$ internal resistive kink, \cite {men18}
where a transition between a $P_r^{-1/3}$ dependence (for moderate values) to a $P_r^{-5/6}$ law is obtained.
A more complete linear study over a larger range of ($\eta$,  $P_r$) values is beyond the scope of the present study.

Note that, only a moderate spatial resolution with $h_{max} = 0.03a$ (for the imposed maximum edge size triangle)
and a relatively large time step $\Delta t \simeq 0.01 t_A$ are sufficient for a good convergence of the results for this linear phase. 
The corresponding total number of triangles $n_t$ remains relatively low with $n_t \simeq 20000$ because a strong automatic mesh adaptation
is not required when the current sheet is not yet formed.
The reader can also refer to Appendix C to see the numerical convergence with the spatial discretization.

\subsection{Non linear phase of the resistive kink, magnetic reconnection process, and growth of
plasmoid chains}

As it is the crucial term in driving the reconnection process, we focus on the time evolution
of the maximum current density $J_{max}$ for resistivity values $\eta$ situated in the approximated range $[10^{-6} - 10^{-2}]$.
The results are plotted in Fig. 7 for a few resistivity values.
First, we note the increasing time delay for triggering the instability as the resistivity is decreasing.
Moreover, the linear phase for kink mode is also longer when the resistivity value is smaller as a consequence
of the decreasing linear growth rate with decreasing resistivity (see above). Consequently, it takes a longer
time to enter the non linear phase and the ensuing reconnection stage for smaller resistivity.
For the runs employing the highest resistivity values, typically for $\eta \simgt 2 \times 10^{-4}$, the behavior
proceeds like the typical case described above with a peak of the maximum current density $J_{max}$ that is an
increasing function of the resistivity inverse.
For $\eta \simlt 2 \times 10^{-4}$, the behavior of $J_{max}$ exhibits additional oscillations superposed to the
time increase, which are due to the presence of plasmoids (see below and in Appendix A). 

The growth of plasmoids obtained in the context of ideal MHD instabilities (tilt and coalescence
setups) can be characterized by an second increased slope in the semi-log plot of the maximum current density evolution,
before reaching an oscillating behavior with a constant time-average value (see also in Appendix A).  \cite {hua17, bat20a, bat20c}
The measure of this slope has thus been used in order to estimate the maximum linear growth rate of the plasmoids. Following the same
procedure in this study (see the fitted exponential laws for the three lowest resistivity runs in
Fig. 7), leads to a value $\gamma_p t_A \simeq 0.45$ that is nearly independent of the resistivity,  where  $\gamma_p$ represents an
instantaneous plasmoid growth rate. When using the appropriate normalization with $\tau_A = L/V_A$
instead of $t_A$, where $L$ is the half-length of the current sheet and $V_A$ is the Alfv\'en velocity
during reconnection based on the upstream magnetic field $B_u$, a value of $\gamma_p \tau_A \simeq 0.25$ can
be deduced. We estimate $L \simeq 1$ and $B_u \simeq 2$ during reconnection. 
Indeed, the magnetic field is observed to undergo an amplification by a factor of order two
during the formation of the current sheet, in a similar way as previously reported for tilt instability setup (see
Fig. 6 in Baty et al.).  \cite {bat20a}. The growth of the plasmoids for the resistive kink is consequently slightly sub-Alfv\'enic,
contrary to the super-Alfv\'enic growth in tilt/coalecence setups where values $\gamma_p \tau_A \simeq 10-20$ were reported.  \cite {hua17, bat20a, bat20b}

As concerns the point of the estimate of the reconnection rate, the use of the maximum current density
is more problematic compared to the diagnostic used for the tilt mode (see Appendix A). Indeed,
the reconnection for the resistive kink proceeds at an essentially non constant  (time varying) $J_{max}$.
Thus, another diagnostic must be taken that is better suited to this case. For typical resistive modes like kink/tearing instabilities, a possible
approach consists in measuring the width of the $m = 1$ island as a function of time, as it
contains the reconnected field lines. \cite {bisk09, white91}
We preferred in this study another equivalent method
consisting in evaluating the remaining flux to be reconnected as a function of time. This is done
by measuring the off-axis displacement of the magnetic axis $\xi_0 (t)$ from the original geometric centre.  \cite {bat91}
This is illustrated in Fig. 8 with a plot of $\xi_0 (t)/a$ versus time for two runs using
different values of the resistivity. 
Indeed, for relatively large values of the displacement, e.g. for typically $\xi_0 (t)/a  \simgt 0.2$, the time increase is due to
the reconnection process. Consequently, its corresponding time derivative can be easily deduced to 
estimate some related reconnection rate $d \xi_0/dt$, as done in Fig. 9.
First, one must note that the SP/Kadomtsev reconnection rate following the usual $\eta^{1/2}$ scaling is recovered for
$\eta \simgt 2 \times 10^{-4}$. However, a transition towards another scaling law occurs for smaller resistivity values.
Consequently, the critical resistivity value of $2 \times 10^{-4}$ for the transition between the two regimes
corresponds to a critical Lundquist number $S_c \simeq 10^4$ for our resistive kink setup with $P_r = 1$.
A minimum value of $d \xi_0/dt \simeq 0.08$ for our reconnection rate is obtained close to the transition, whilst the rate seems to be even slightly
higher for our lowest resistivity values. The above minimum value of the reconnection rate can be translated into a normalized value
$ \frac {1}  {V_A B_u}  \frac {d\psi}  {dt}  \sim 0.02$, as the value of the displacement $\xi_0$ is
of the order of the flux value $\psi$ in our units.
The latter result confirm the acceleration of the reconnection speed with respect to the SP regime
when the resistivity is low enough. This second regime of reconnection corresponds to the presence of
plasmoids (see Fig. 10). 
Moreover, when the resistivity is low enough, 
typically for $\eta = 7.7 \times 10^{-6}$ in panel (c), the system behavior exhibits a current structure with Petschek-type structure that is very similar to results
obtained for tilt setup. \cite {bat20c}. These Petschek-like shocks appear to be even more developped
for $\eta = 3.9 \times 10^{-6}$ in panel (d). The latter appears as a current layer with pairs of slow-mode shocks emanating 
from a small central region containing a few plasmoids.
The occurrence of the latter dynamical Petschek solution was
shown to be related to coalescence events between primary plasmoids leading to the formation of monster plasmoids. \cite {bat20c}
This tendency to develop a time dependent Petschek-like structure has been also reported using numerical
experiments in the full MHD framework, with the aim to study eruptive events in the solar corona or heating mechanisms in the low
solar atmosphere. \cite {mei12, ni16} 

The ability of our numerical scheme to capture the associated small-scale current features is illustrated in Appendix B, 
and a convergence study can also be found in Appendix C.

\section{Discussion and Conclusion}
\label{discussion_conclusion}

In this study we have studied the effect of the forming current sheets mechanism via the 2D resistive kink instability on the
onset of plasmoid chains formation and ensuing fast magnetic reconnection regime. To this aim, we have considered initial
configurations with a core reversal current density that are resistively unstable, inspired by the ``current hole" problem in tokamaks. 

First, we have checked the linear properties of this 2D $m = 1$ resistive kink mode. The results are very similar
to the well known $m / n = 1/1$ internal kink mode in 3D cylindrical geometry, with linear growth rate scaling as $\eta^{1/3}$.
Second, the non linear evolution leading to the formation of a current sheet and to the associated reconnection
is investigated for a relatively large range of different resistivity values.
When the resistivity is large enough, i.e. corresponding to a local Lundquist number $S  \simlt S_c$ (with  $S_c = 10^4$ for $P_r = 1$),
the classic Sweet-Parker (or equivalently Kadomstev) reconnection regime is recovered with a reconnection
rate scaling as $\eta^{1/2}$. Alternatively, when $S \simgt S_c$, the forming current sheet is itself unstable leading to the
presence of plasmoids during the ensuing magnetic reconnection. The time scale involved for the growth of plasmoids
is estimated to be sub-Alfv\'enic with a growth rate $\gamma_p \tau_A \simeq 0.25$, that is moreover quasi-independent
of the resistivity (and corresponding Lundquist number).
This is one order of magnitude lower when compared to results obtained using ideal MHD instability setups, where
$\gamma_p \tau_A \simeq 10-20$ have been reported for coalescence and tilt mode. \cite {hua17, bat20a, bat20b}
Moreover, in the latter cases a non power scaling law of $\gamma_p $ with $S$ was obtained in agreement
with Comiso et al. theory. \cite {com16, com17} In the present study, the plasmoid growth rate is in rough agreement with
a value $\gamma_p \tau_A \simeq 0.4$ deduced from the alternative model of ``ideal" tearing mode
(for Harris-type current sheet with $P_r = 1$). \cite {puc14, puc18}
The factor of two lower may come from the effect of the magnetic reconnection flow (not included in the two models), which
is known to lead to a substantial stabilizing effect. \cite {tol18, hua19}
 
Thus, we infer that, when the mechanism at the origin of the current sheet formation is due to the development of
an ideal  MHD instability (so the corresponding time scale is Alfv\'enic), the model proposed by Comisso et al. applies. \cite {com16, com17}
Conversely, if the current sheet formation is driven by a resistive instability (i.e. on a slower time scale
depending on the resistivity), the model developped by Pucci and coworkers (e.g. ``ideal" tearing mode) is suited to predict the plasmoids behavior. \cite {puc14, puc18}
In this latter case, the plasmoid growth rate is at most Alfv\'enic in contrast to the ideal setups for which
a super-Alfv\'enic growth is allowed. This is not completely surprising as the two models differ in the assumptions taken
for the current sheet formation. Note that it is however possible to also recover such sub-Alfv\'enic growth rate when
taking for the functional form of the current sheet formation rate a proper value in the model of Comisso et al. \cite {com20}
Consequently, more work is probably needed to highlight this point.

In this work, we have also emphasized the resemblances/differences between tilt and resistive kink setups on the ensuing
reconnection stage in the high Lundquist number regime. The obvious difference (that also holds for SP regime) is 
that magnetic reconnection essentially proceeds at a roughly constant maximum current density for tilt instability.
Conversely, for resistive kink instability, the current density in the current sheet is continuously increasing in time during reconnection.
Using the appropriate
diagnostic to estimate the reconnection rate, lead to a normalized reconnection rate of order $0.02$ in this study, that is
similar (in order of magnitude) to the $0.014$ value reported for tilt setup. \cite {bat20a,bat20b,bat20c}
This is also in agreement
with values generally quoted in the literature. Finally, for the lowest resistivity values employed in kink/tilt studies, our
simulations have revealed current structures typical of Petschek-type reconnection. Future work at even
lower resistivity values is needed to investigate the latter point, in particular with the scope to reach the huge Lundquist
values typical of the plasma parameters of the solar corona.
In the present work, we have restricted ourselves to cases with Prandtl number equal to unity for
the nonlinear evolution. In a future work, it would be thus interesting to explore the effect of the viscosity with our code. Indeed,
the plasmoid instability is believed to drastically influence the MHD turbulence through the associated fast reconnection achieved
in this way. \cite {dong18} It would be also interesting to explore three-dimensional effect.
In the late years, considerable progress has been achieved in numerical simulations mainly due to
the speeding up of computers. Consequently, the plasmoid instability regime has become accessible to three-dimensional MHD
studies, showing that  an enhanced reconnection regime is able to set in already at $S \simeq 10^3$ (see Ref. 38). 

The Lundquist number reached in this study is high enough in order match the relevant values for
tokamaks. Indeed, the relevant $S$ value for the internal disruption associated with the internal kink mode is
$S  \simeq 10^5$, as $S = 0.004 S^*$ ($S^* = 2.5$ $ 10^7$ being a standard Lundquist number value defined
in terms of the toroidal magnetic field)  \cite {gun15}. The corresponding width of the Sweet-Parker current layer is thus estimated to be $a \simeq 1$ cm,
and the smallest length scale associated to the plasmoid structure is probably of order $1$ mm or even smaller, reaching
thus a scale close to the the kinetic ones. Kinetic effects could be incorporated to our model in order to address this
point. For example, the plasmoid instability has been shown to facilitate the transition to a Hall reconnection 
in Hall magnetohydrodynamical framework with an even faster reconnection rate of $\sim 0.1$ \cite {hua11}.
The smallest length scale associated to the plasmoid structure for $S = 10^6$ remains larger than the kinetic scale 
that is of order $10$ m, when considering a solar loop structure and taking a length $L = 10^7$ m. However, as very high
Lundquist number (at least $10^{10}$) is required for the solar corona, kinetic effects could also play a role if the kinetic scale
is reached via the plasmoid cascade at such huge Lundquist number.  \newline

\textbf{ACKNOWLEGDMENTS}\newline
The author wishes to acknowledge Luca Comisso for fruitful discussions. It also a great pleasure to thank one
anonymous referee who helped to ameliorate the content of the manuscript, and who
recognized in this work the 'high quality of the numerical method and the high accuracy of the simulations, which is much higher than other works on
plasmoids formation in several instability or current sheet setups'.
Unfortunately, a strong and incomprehensible opposition by another
reviewer has prevented publication in a peer reviewed journal, with some arguments mostly based on belief (that plasmoid phenomena are just
a numerical artefact due to insufficient numerical resolution, and ignoring all the many published works on the subject)
and not on the careful reading of the manuscript. Finally, the measure of the rate of plasmoid growth (using the maximum
current density amplitude as the main diagnostic) obtained in this work was also strongly criticized (this is apparently a very sensitive controversial point 
to discriminate between the different theoretical models), while this point has been previously addressed in detail using other MHD setups
(see Refs. 12-14).

\appendix
\section{Resistive kink versus ideal tilt}

 It is instructive to compare the typical time evolution of the current density obtained for the ideal tilt instability (used in previous studies)  \cite {baty19, bat20a, bat20b, bat20c}
 to the evolution of the resistive kink instability investigated in this work. Indeed, the maximum current density $J_{max}$ is plotted as the function of time in
 Fig. 11 for three different resistivity values and a fixed magnetic Prandtl value $P_m = 1$. Note that, the term preventing the equilibrium diffusion is
 not implemented for the tilt setup, as the early evolution stage is relatively fast and independent of the resistivity.

Panel (a) corresponds to the Sweet-Parker regime as the resistivity is relatively high (i.e. $\eta = 2 \times 10^{-3} $), and it clearly shows an important first distinction in the
magnetic reconnection process between the two setups. Indeed, reconnection of magnetic field lines driven by the resistive kink instability essentially proceeds during the increase
in time of the current density and ends up when the maximum is reached. This is not the case for the tilt instability, as the reconnection phase is triggered at a time close
to the peak and it subsequently proceeds at a roughly constant maximum current density. 
As a consequence, the diagnostic using $\eta J_{max}$ to estimate the reconnection rate and to compare to the steady-state Sweet-Parker model is fully justified for tilt case.
For the resistive kink instability, the use of the growth of the $m = 1$ magnetic island is a good alternative that is commonly used. \cite {bisk09}
Another option is to follow the progression in time of the initial magnetic axis displacement, as it will corresponds to the last magnetic field lines to be reconnected
(see main text). \cite {bat91}

Panel (b) corresponds to a resistivity value $\eta = 1.25 \times 10^{-4}$, for which a few plasmoids are observed to grow in the two setups.
First, one must note the increased delay in the current increase compared to previous panel, mainly as the consequence of
the resistivity dependence of the kink mode linear growth rate. The oscillations seen on the current density plots correspond to the effect of
the plasmoids, thus leading to the obtention of a stochastic reconnection regime.

Panels (c)-(d) correspond to the plasmoid-dominated regime, as the resistivity value is very low, e.g. $\eta = 7.7 \times 10^{-6}$.
First, the increased time delay is even higher compared to previous panel. Second, one can also see an obvious additional difference between the two setups.
Indeed, an abrupt change of slope just before reaching a peak amplitude 
is evident for the tilt case (see panel (c)) that has been shown to be the consequence of the early growth of plasmoids. \cite {baty19, bat20a, bat20b}
The subsequent reconnection phase is stochastic with a current density oscillating around a constant time-averaged value. 
For the resistive kink instability (panel (d), there is a less spectacular change of slope, which is simply continued with superposed
oscillations during magnetic reconnection phase. This second slope fitting the exponential time increase of the maximum current density in
presence of plasmoids is fundamental in order to evaluate an instantaneous plasmoid growth rate $\gamma_p$ (see main text).

\section{Small-scale current structures}

The ability of our adaptive mesh procedure to capture all the small-scale features of the current structure is illustrated in Fig. 12.
Indeed, we zoom-in on a portion of the current layer in order to point out a few plasmoids. Interestingly, a coalescence event
between the two upper plasmoids is visible in panel (a). It occurs via the small-scale current sheet of negative value. that
is perpendicular to the main current layer (with positive current density). An even bigger zoom-in on this event overlaid with
the adapted mesh structure is also plotted in panel (b). This clearly shows the very high number of generated triangles to capture the finest
structures (the positive/negative current layers). Typically, at any time our scheme is able to cover the current structures
with a few tens of elements with a minimum reached edge size $h_{min} \simeq 10^{-5}$, the imposed maximum edge size value
being $h_{max} \simeq 10^{-2}$. However, the corresponding total number of triangles $n_t$ used for the lowest resistivity case remains
lower than the maximum number allowed of $\sim 10^6$.

\section{ Mesh convergence} 

In Appendix B, it is shown that, at any time, many small triangles are generated via the adaptive mesh procedure in order to capture the smallest
structures with a few tens of elements. This is due to the fact that, the maximum allowed number of triangles (that is imposed during the runs) is never reached.
Additionally, in this Appendix, we carry out additional simulations for two cases, in order to demonstrate the convergence of our results for
SP regime with $\eta = 2 \times 10^{-3}$ (see Fig. 3) and for plasmoid regime with $\eta = 3.1 \times 10^{-5}$ (see Fig. 7).
First, we have investigated how the choice of the maximum edge size (of the triangles) $h_{max}$ influences the solution. The results
are plotted in Fig. 13 (a) for  $\eta = 2 \times 10^{-3}$ case. Indeed, reducing the base resolution from $h_{max} = 0.2$ to $h_{max} = 0.25$
has a negligible effect on the whole time evolution of the maximum current density and maximum vorticity. This corresponds to a reduction of order two
for the initial number of triangles. Second, we have investigated the effect of limiting the minimum edge size $h_{min}$ on the same resistivity case.
Indeed, employing values of $h_{min} \simeq 6 \times 10^{-3}$ and $h_{min} \simeq 1.5 \times 10^{-2}$ only slightly modify the converged solution of Fig. 3 where $h_{min} \simeq 3 \times 10^{-3}$
is reached (only the current density is plotted for simplicity), as can be seen in Fig. 13 (b).
Finally, we have investigated the effect of limiting the minimum edge size $h_{min}$ for the case involving many plasmoids with $\eta = 3.1 \times 10^{-5}$.
The minimum edge size reached for the converged solution of Fig. 7 is $h_{min} \simeq 3 \times 10^{-5}$. The results for the maximum current density are plotted in Fig. 14,
and show that reducing $h_{min}$ does not change the average slope (that is important to determine the growth rate as explained in the main text) and
neither the maximum value reached before relaxation.
This is true even for $h_{min} \simeq 2 \times 10^{-4}$. However, as expected for such stochastic solution, the precise scenario for oscillations during the
plasmoid growth phase is affected as much as $h_{min}$ is strongly reduced, as one can see in second panel.

In summary, Our results clearly demonstrate the convergence of the two solutions plotted in the main part of the present paper.

\begin{figure}
\centering
 \includegraphics[scale=0.26]{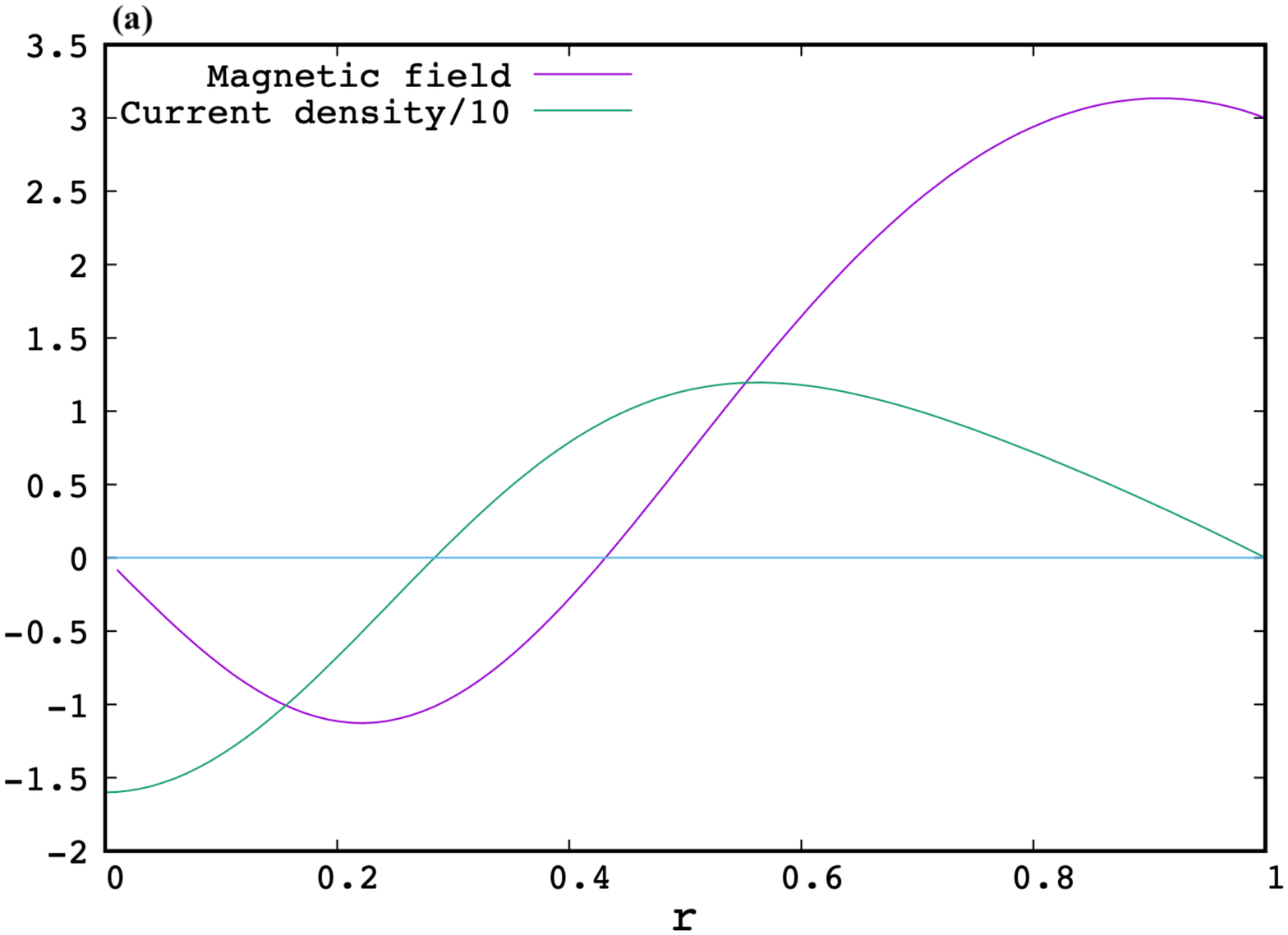}
 \includegraphics[scale=0.28]{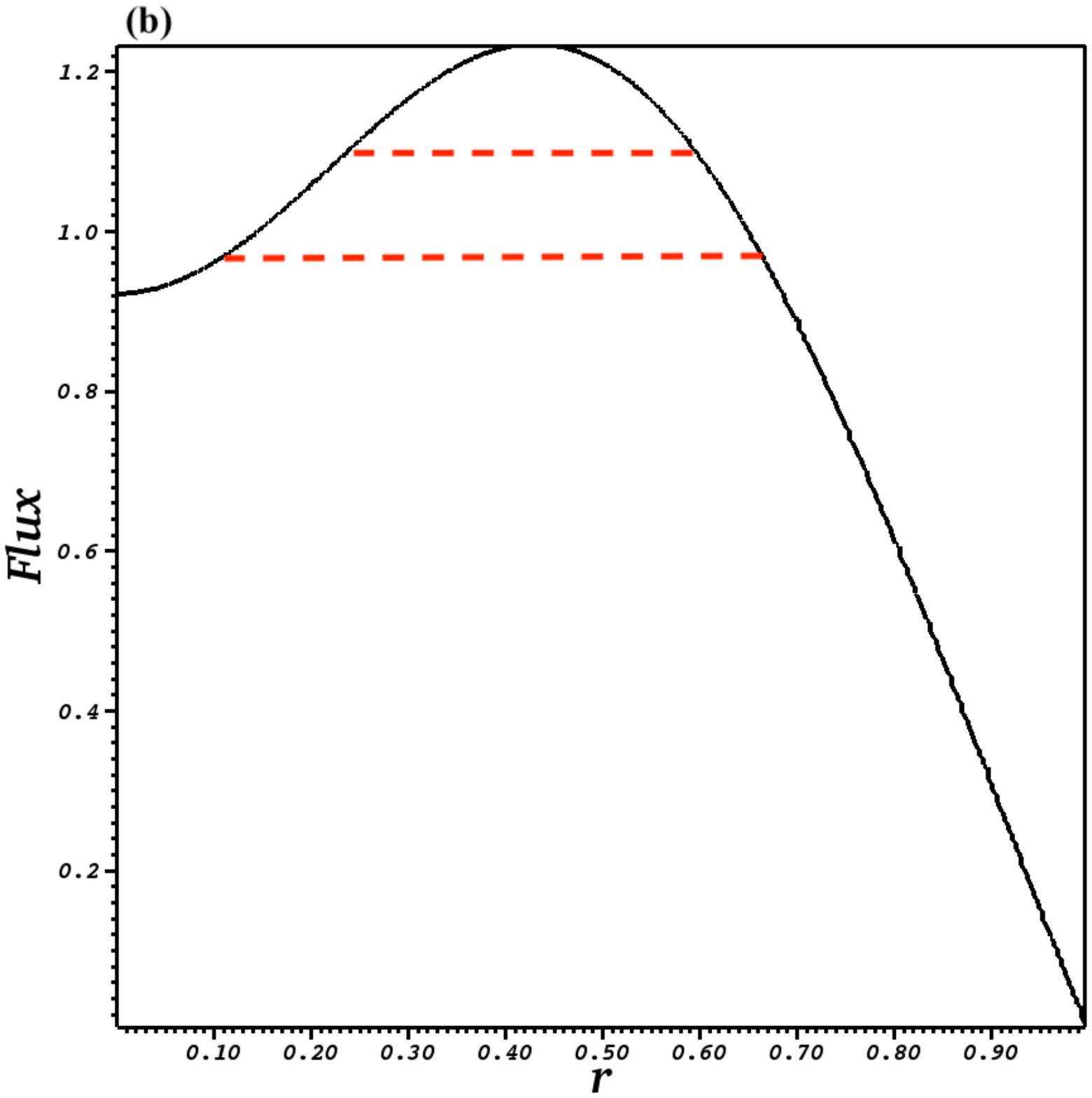}
  \caption{(a) Radial profile of the equilibrium current density $J_e (r)$, and of the corresponding azimuthal
  magnetic field component $B_\theta (r)$. Note that the current is scaled down by a factor of $10$ for better visualization.
  (b) Corresponding magnetic flux profile $\psi_e (r)$,
  where are indicated (see two hatched lines) magnetic field lines to be reconnected on the two sides 
  of the maximum situated at the specific radius $r_s \simeq 0.43a$.
  }
\label{figu1}
\end{figure}

 \begin{figure}
\centering
 \includegraphics[scale=0.20]{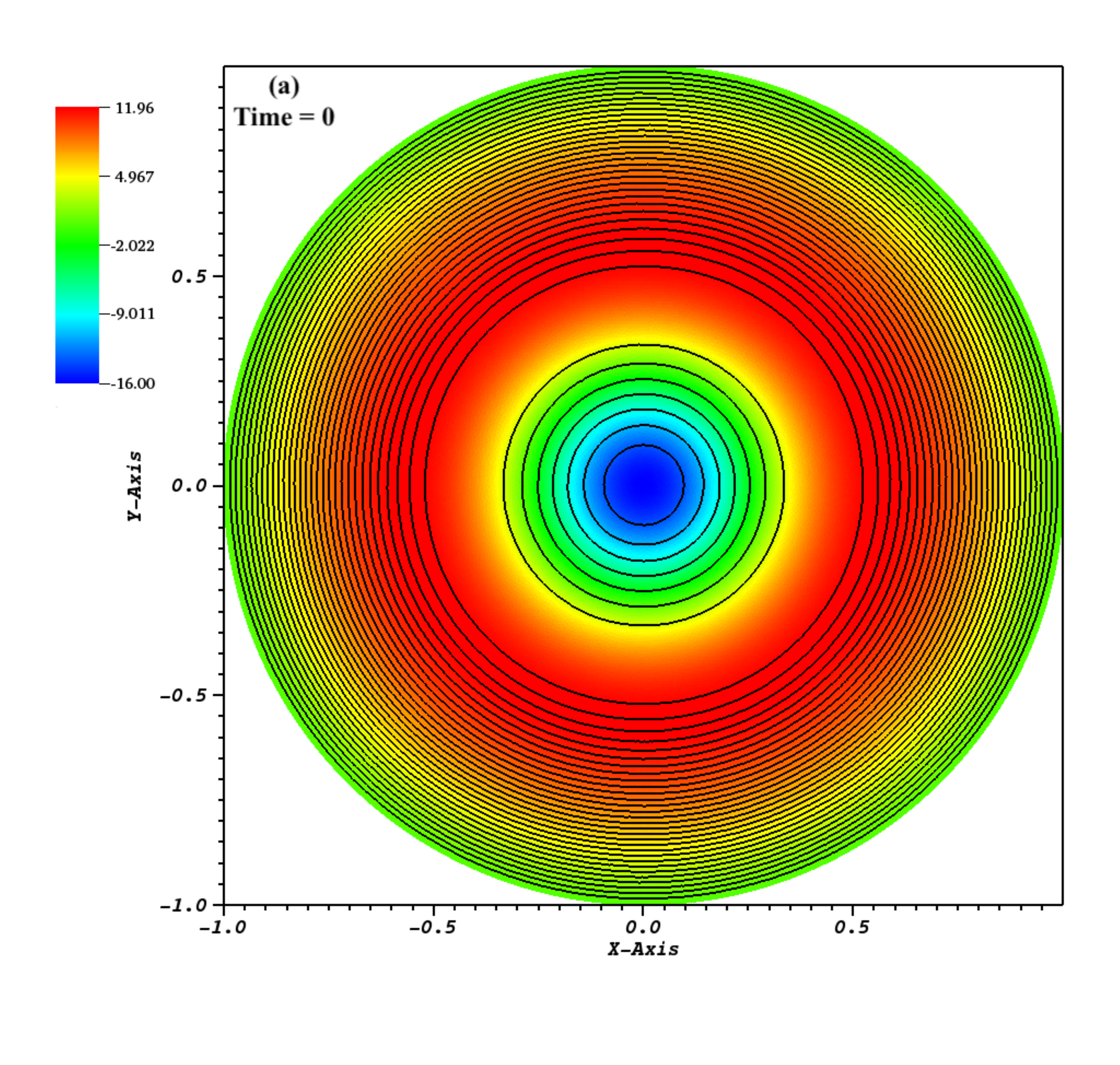}
 \includegraphics[scale=0.20]{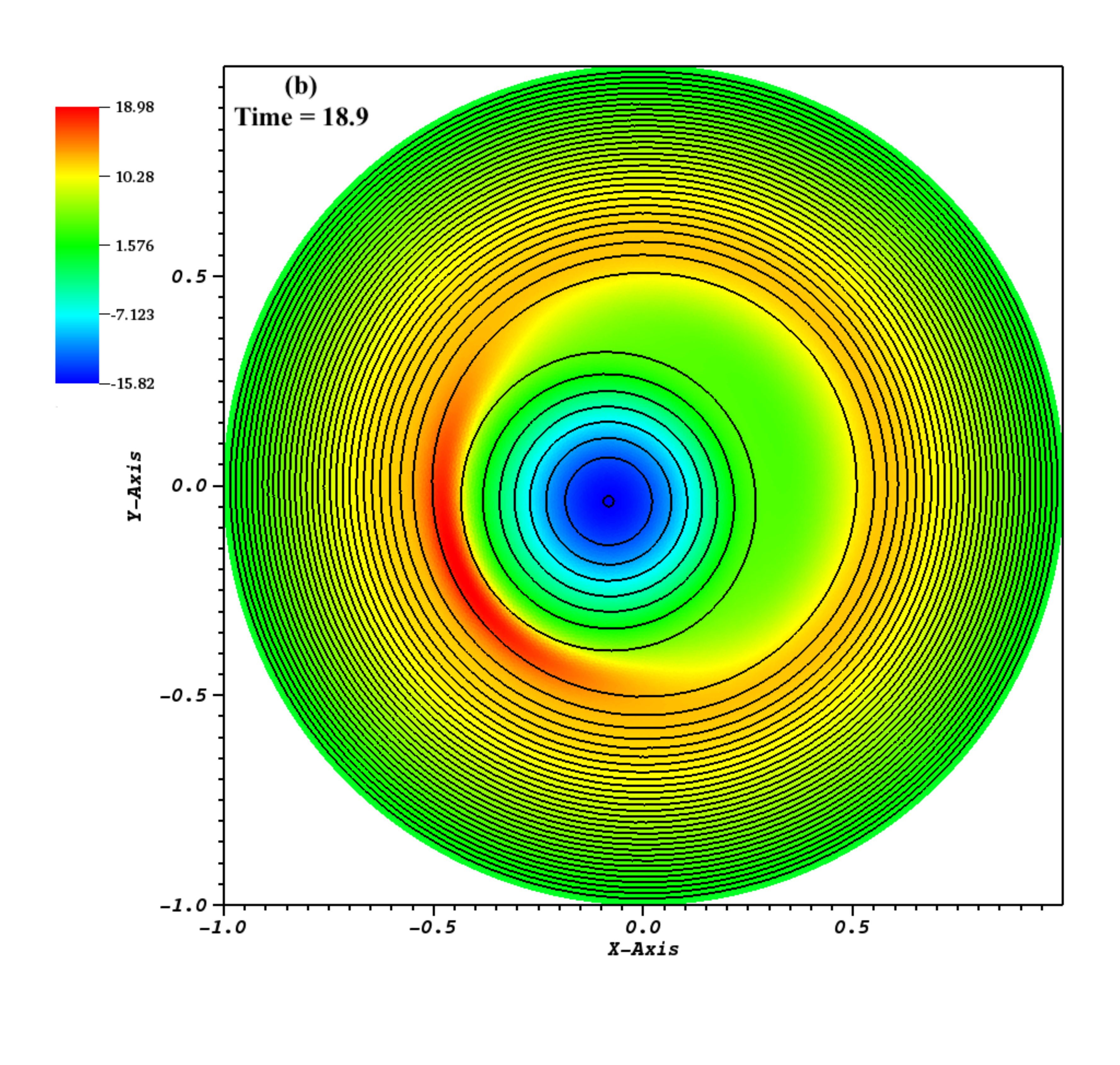}
  \includegraphics[scale=0.20]{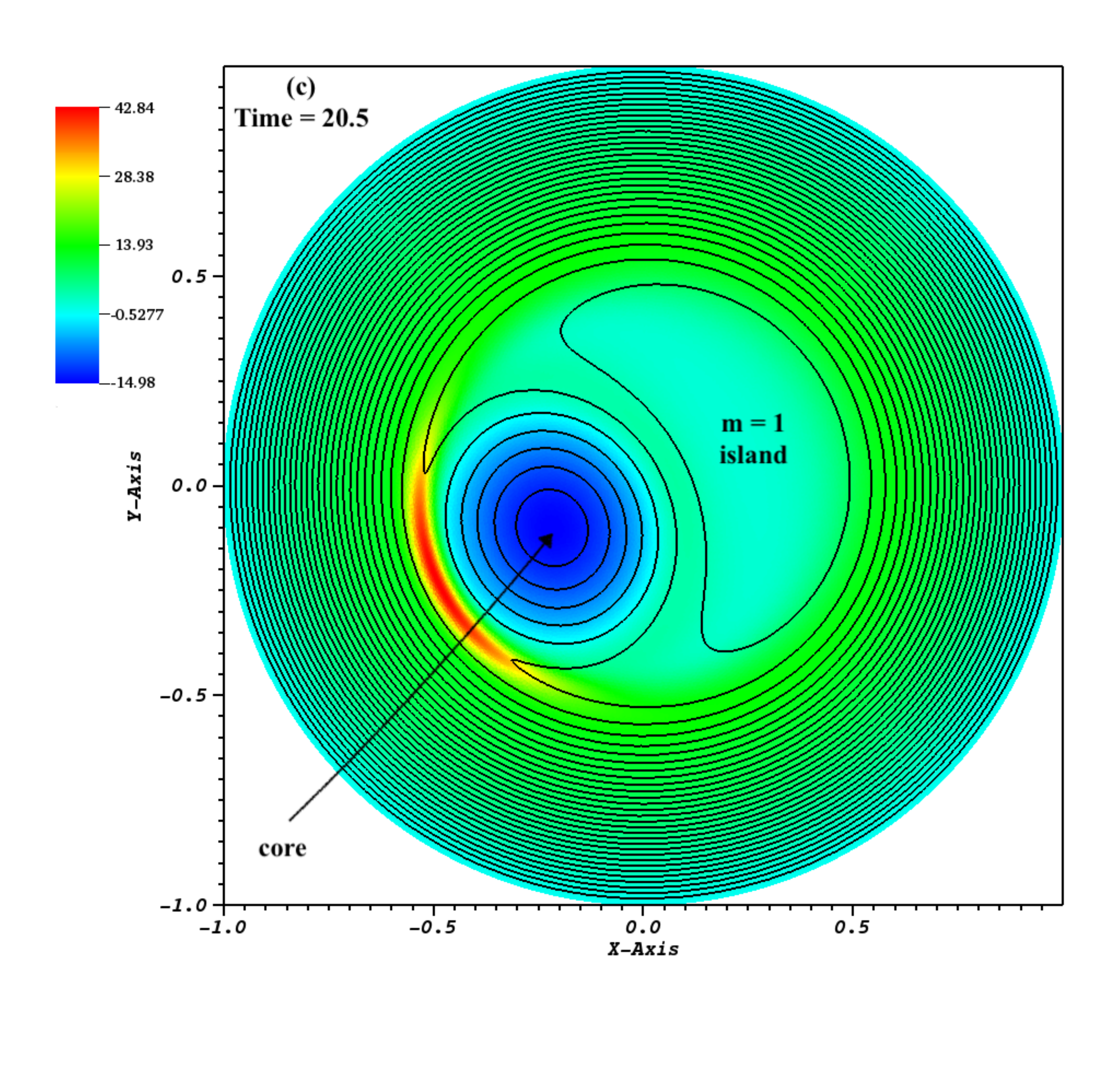}
 \includegraphics[scale=0.20]{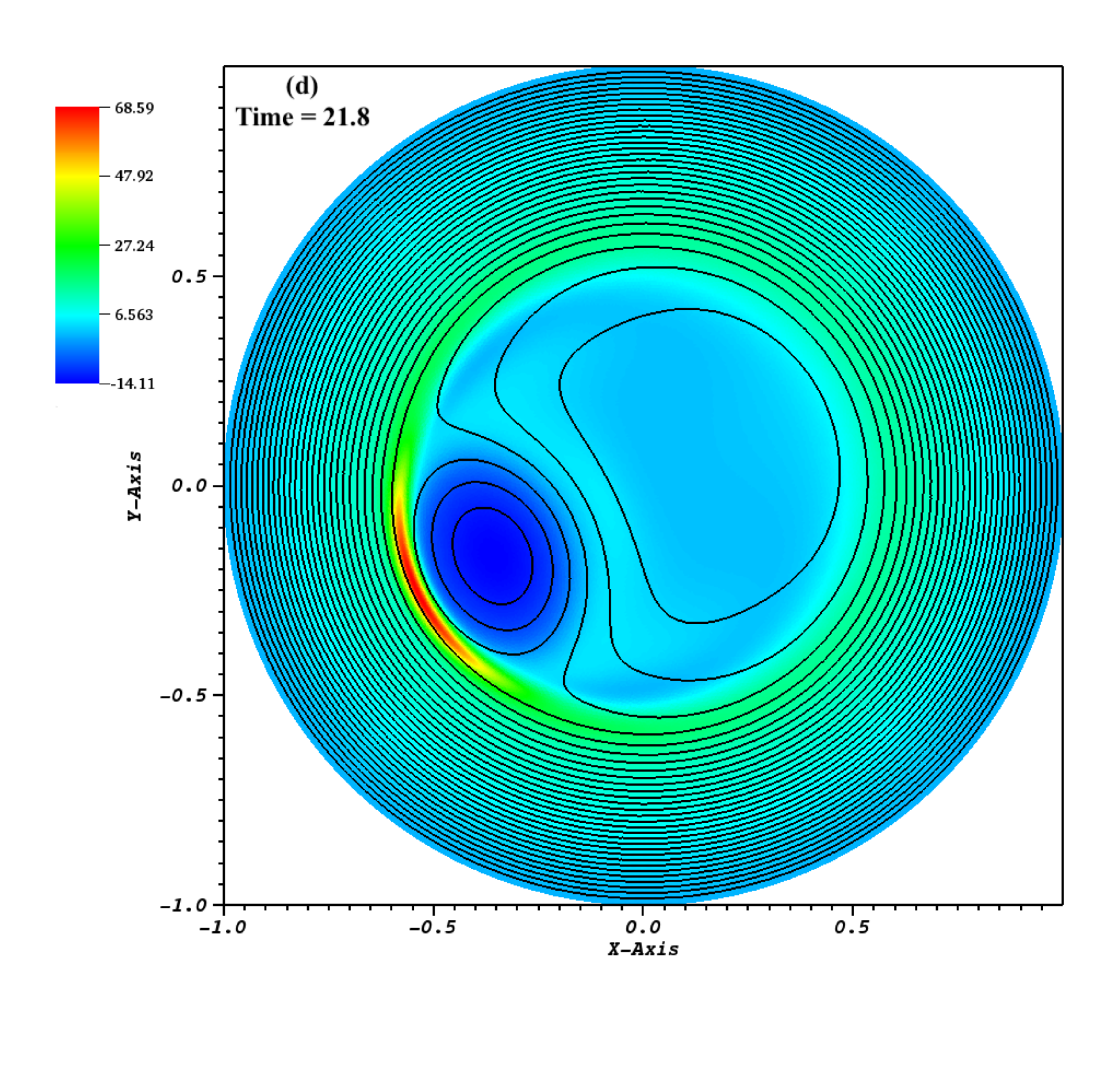}
  \includegraphics[scale=0.20]{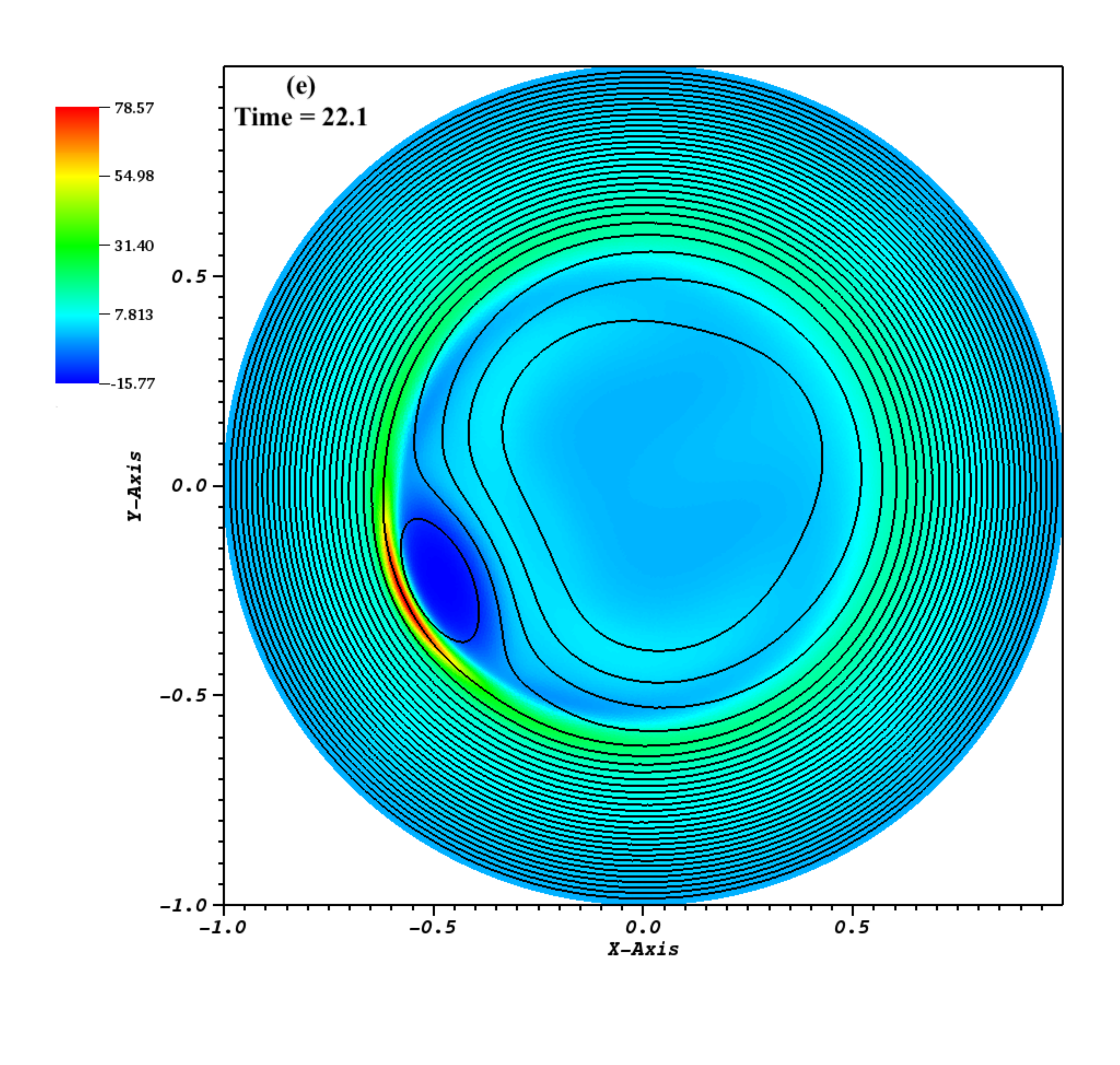}
 \includegraphics[scale=0.20]{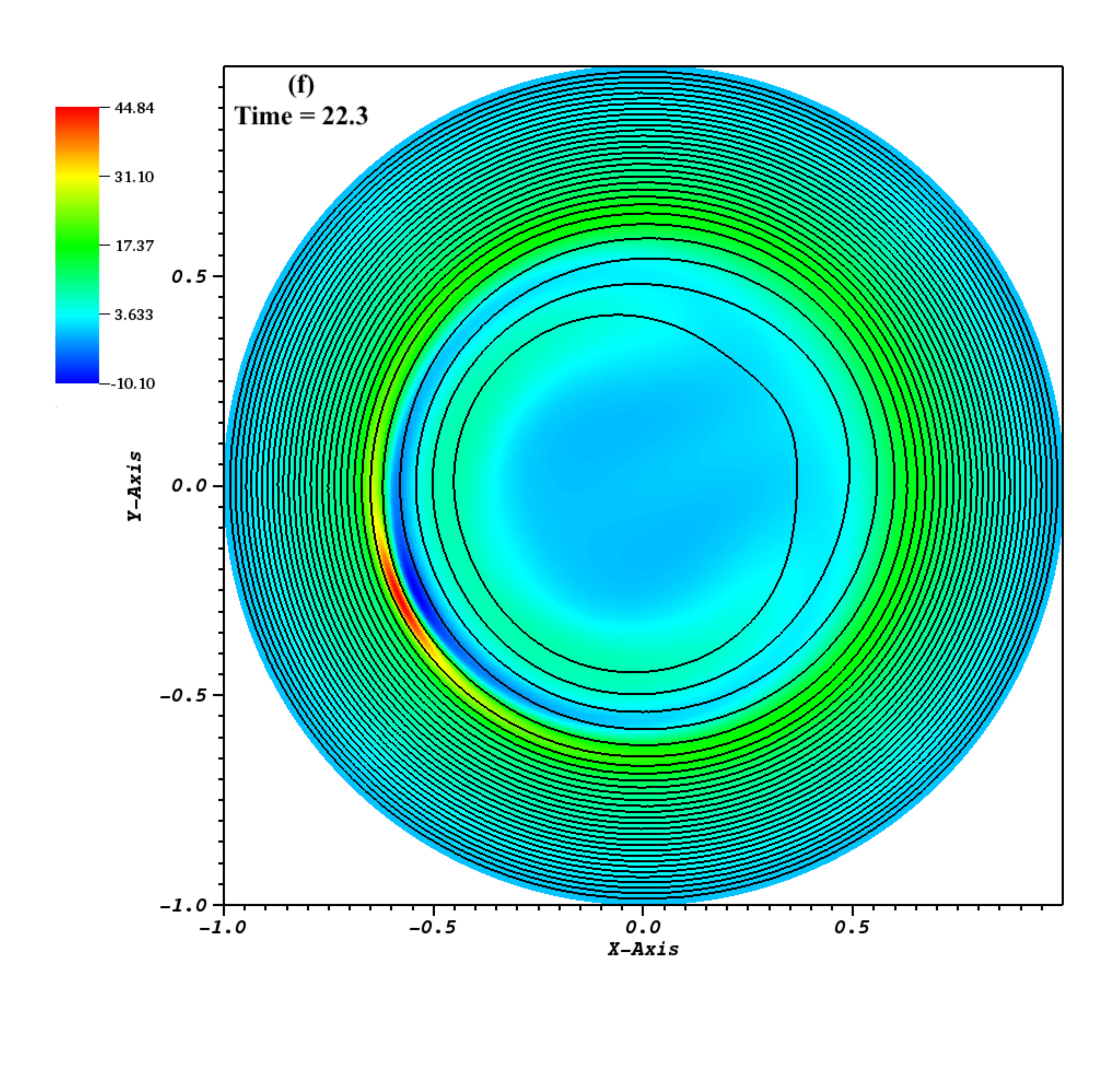}
  \caption{Colored contour maps of the current density overlaid with magnetic field lines taken at different times (see label in units of $t_A$)
  for the run using a resistivity value $\eta = 2 \times 10^{-3}$ and a magnetic Prandtl value $P_r = \eta/ \nu = 1$.
 }
\label{figu2}
\end{figure}

  \begin{figure}[htb]
  \centering
 \includegraphics[scale=0.6]{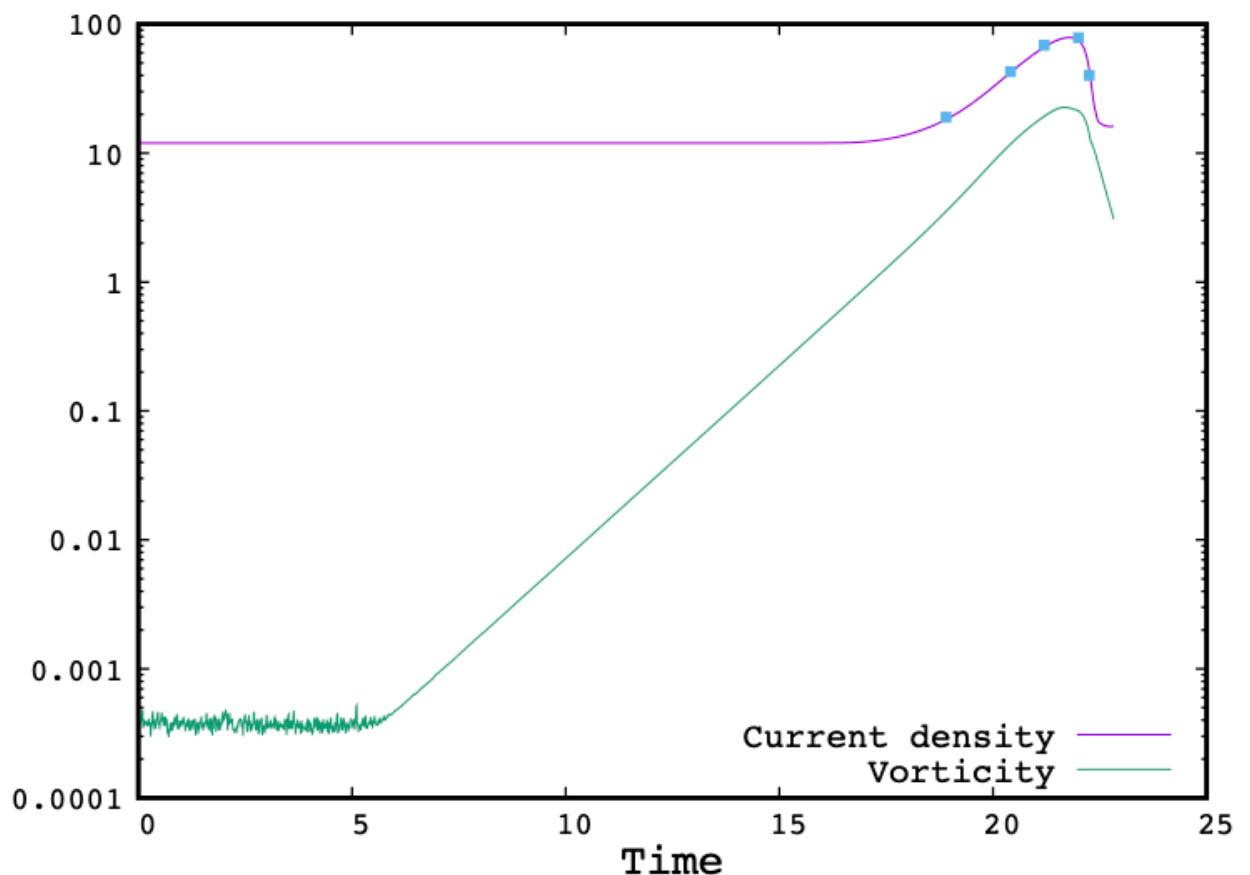}
  \caption{Time history of the maximum curent density and maximum vorticity ($J_{max}$ and $\Omega_{max}$) obtained for the run 
  using a resistivity value $\eta = 2 \times 10^{-3}$ and a magnetic Prandtl value $P_r = 1$. Note that
  $5$ points (using filled squares) are spotted at the times (in units of $t_A$) of the corresponding snapshots showing
  current density in panels (b)-(f) in previous figure.
   }
\label{figu3}
\end{figure}

  \begin{figure}
\centering
 \includegraphics[scale=0.25]{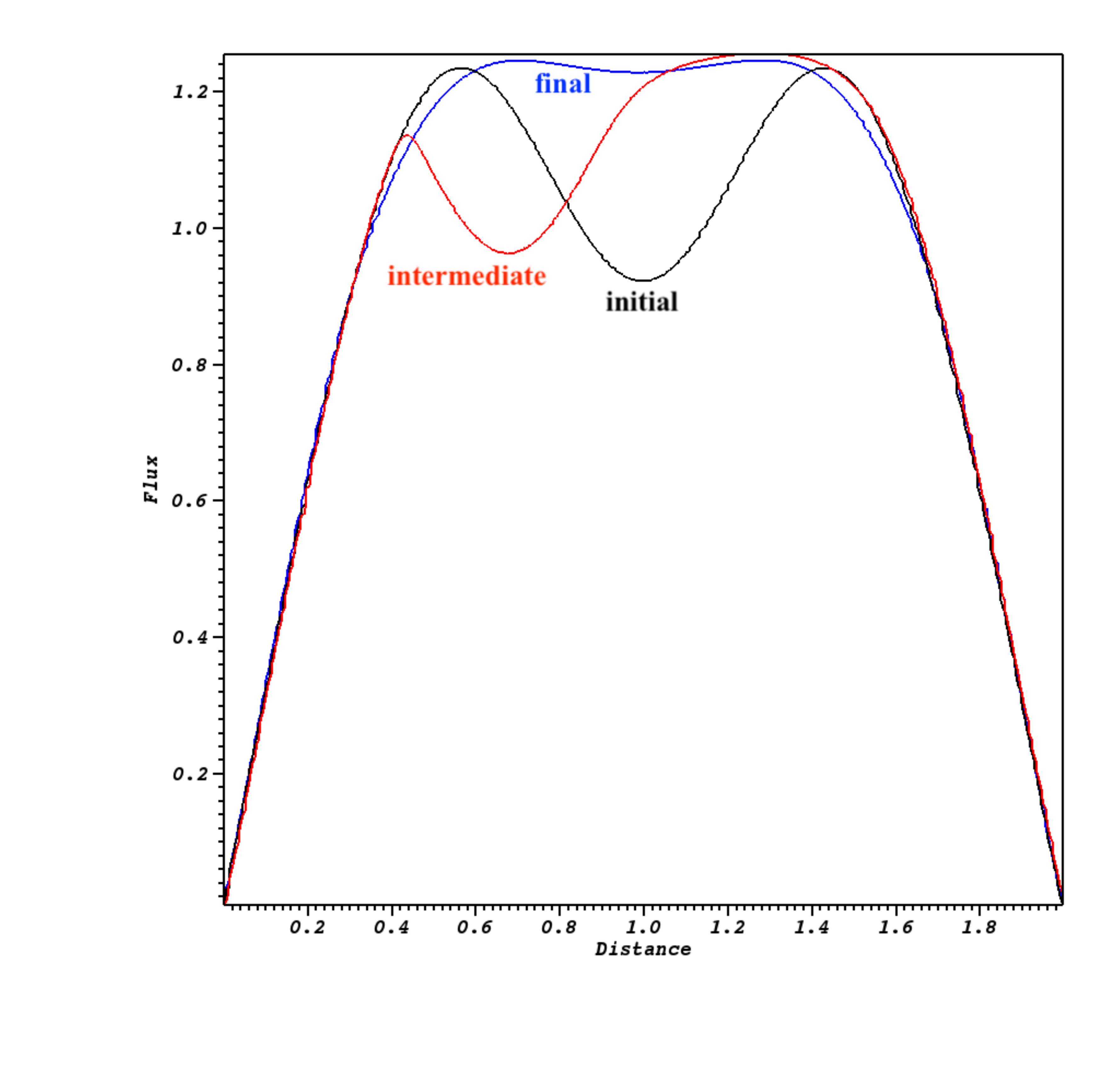}
  \caption{Magnetic flux $\psi$ across the whole diameter (i.e. between $-a $ and $a$) taken at three times,
  for the initial state, during reconnection (intermediate state), and the for final state.
   }
\label{figu4}
\end{figure}

 \begin{figure}
\centering
 \includegraphics[scale=0.6]{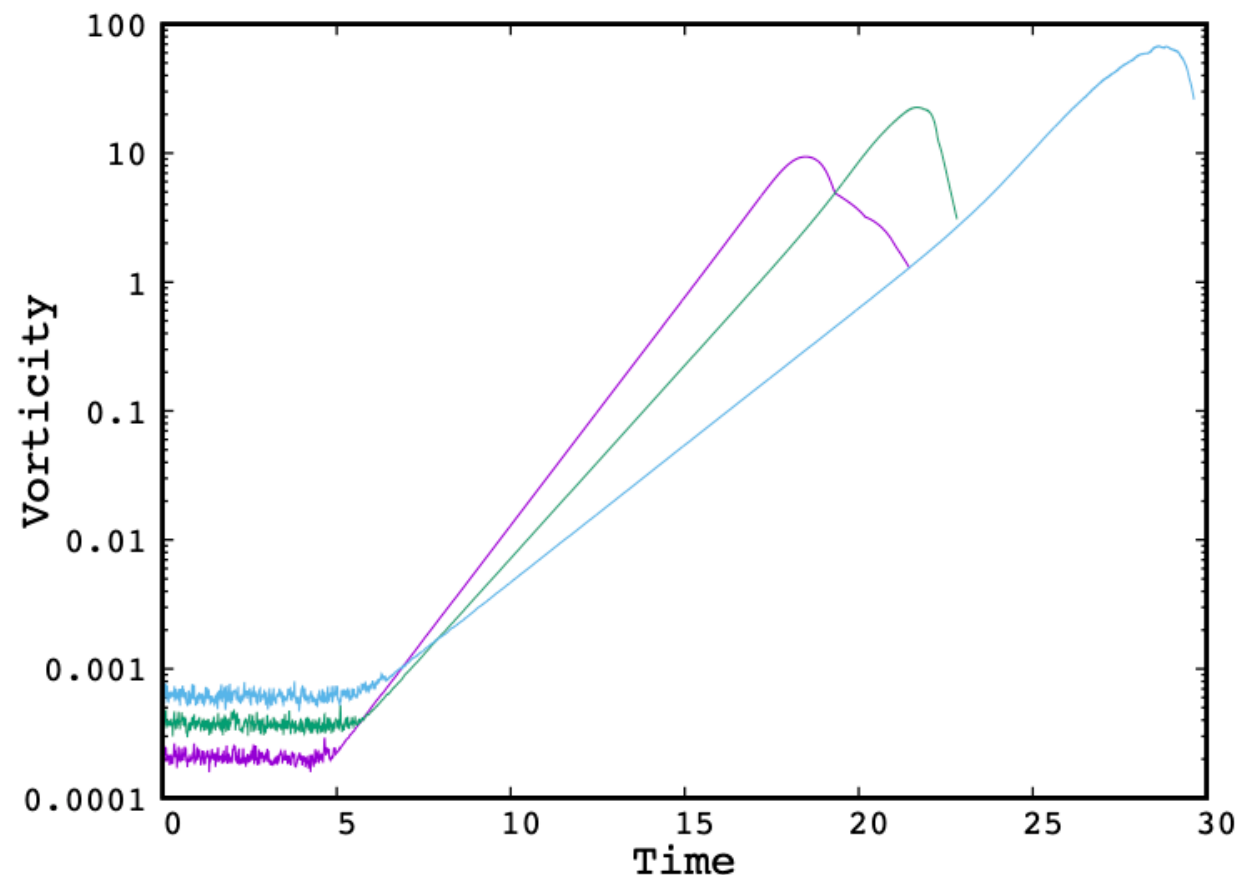}
  \caption{Time history of the maximum vorticity ($\Omega_{max}$) obtained in three runs using
  resistivity values of $\eta =  6 \times 10^{-3}, 2 \times 10^{-3}$, and $5 \times 10^{-4}$ (from leftmost to rightmost curves respectively). A fixed magnetic
  Prandtl number $P_r = 1$ is taken.
    }
\label{figu5}
\end{figure}

  \begin{figure}
\centering
 \includegraphics[scale=0.64]{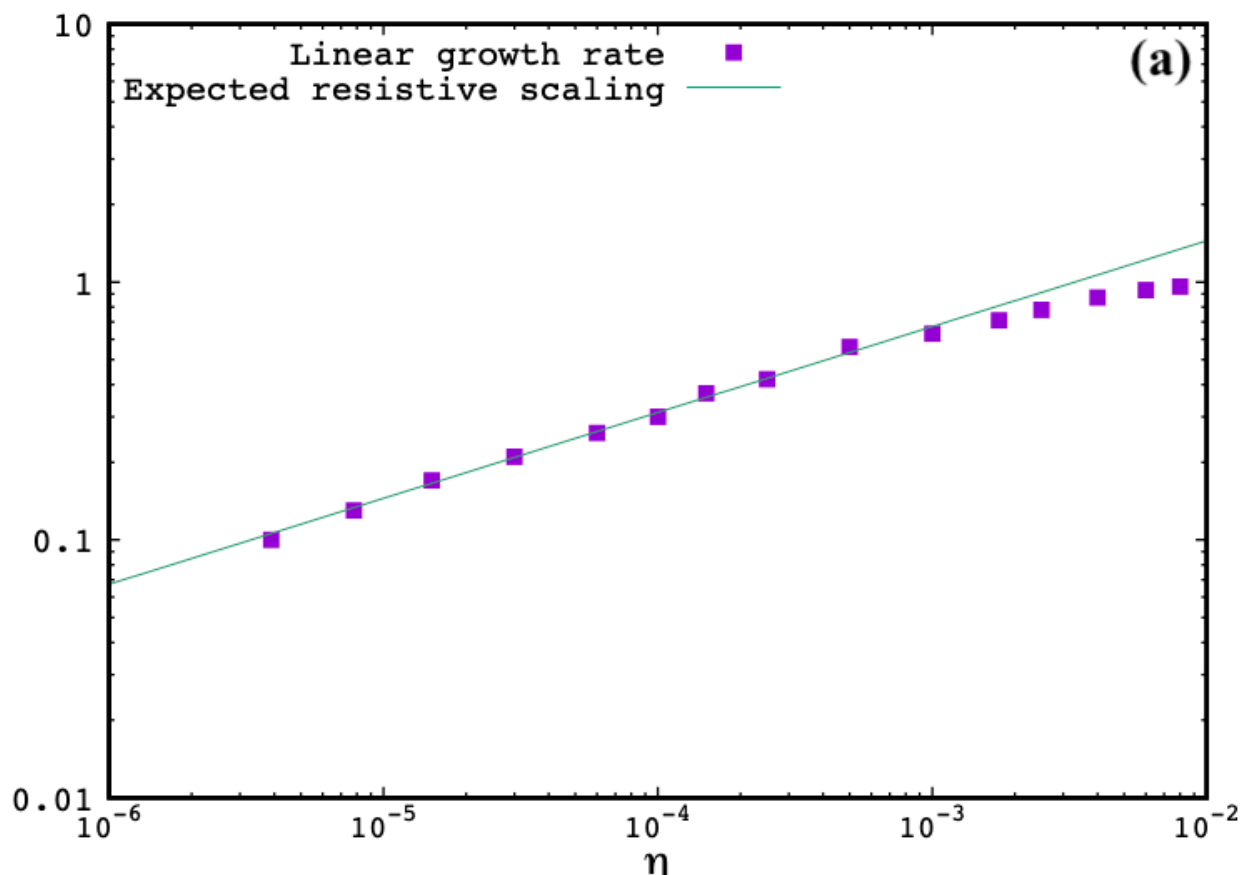}
 \includegraphics[scale=0.64]{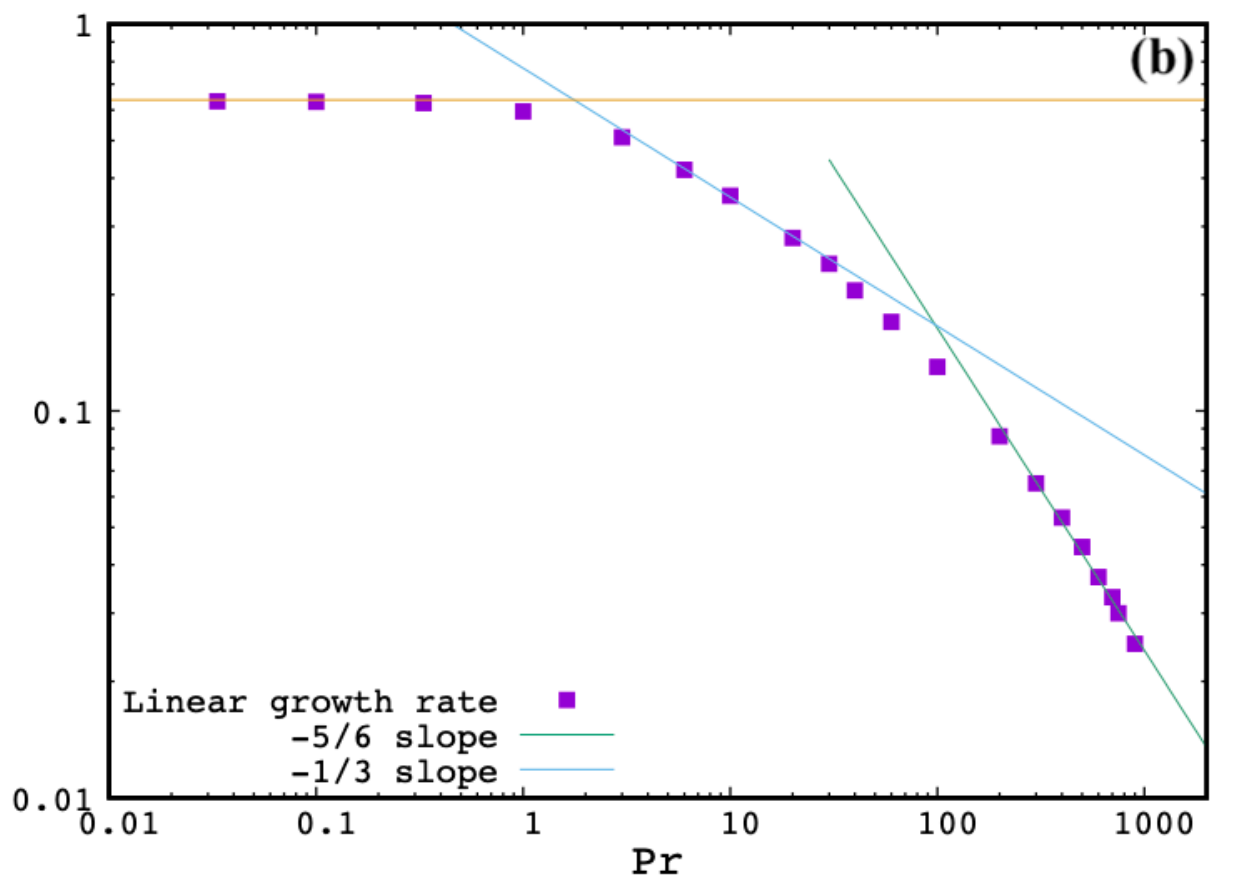}
  \caption{(a) Linear growth rate $\gamma t_A$
  of the resistive kink mode measured from the different runs as a function of the resistivity value $\eta$
  for a fixed Prandtl value $P_r = \nu/\eta = 1$. The expected resistive scaling of $\eta^{1/3}$ is plotted for comparison.
  (b) Linear growth rate $\gamma t_A$ as a function of the Prandl number $P_r$ for a fixed resistivity
  value $\eta = 10^{-3}$. The scalings of $P_r^{-1/3}$ and  $P_r^{-5/6}$ are plotted for comparison (see text).
   }
\label{figu6}
\end{figure}

 \begin{figure}
\centering
 \includegraphics[scale=0.7]{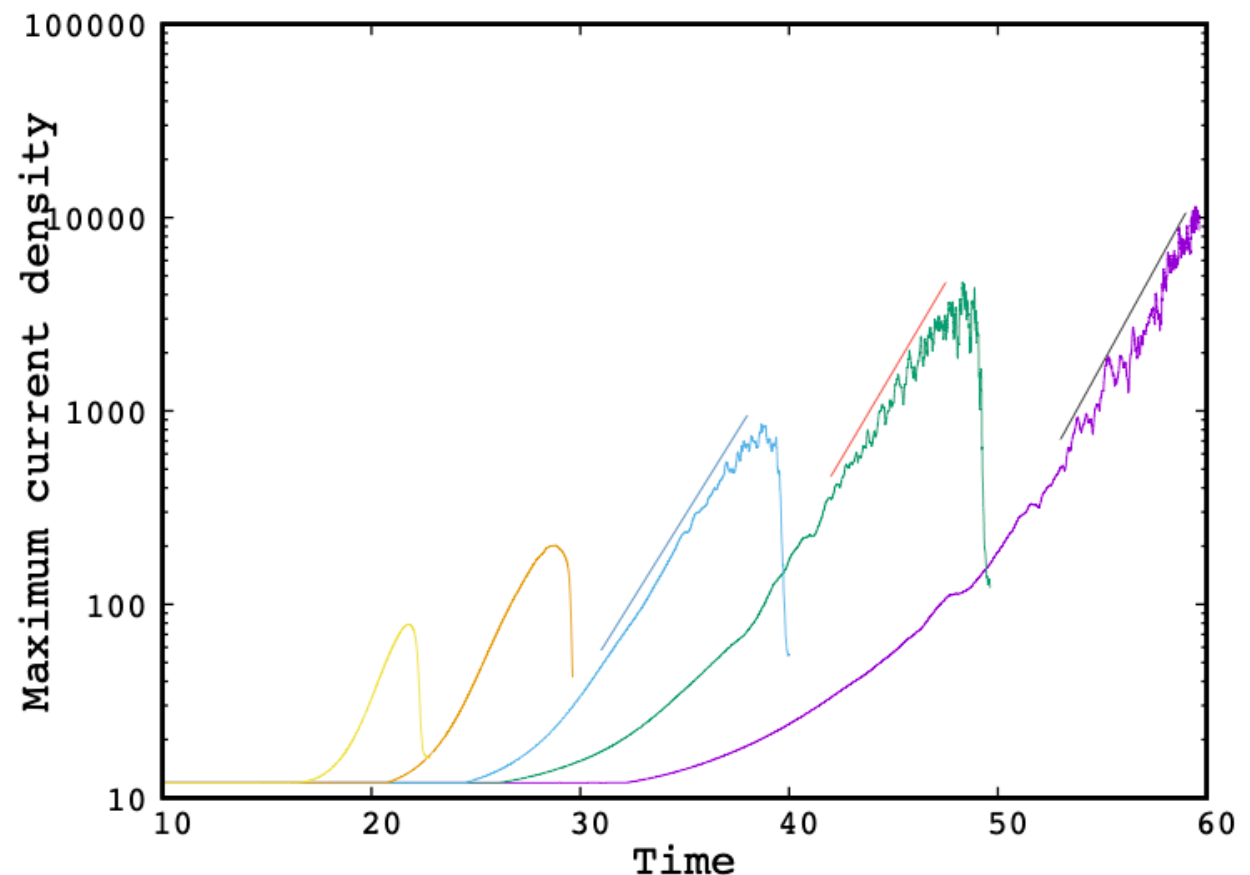}
  \caption{Maximum current density $J_{max}$ as a function of time for five different resistivity values, $\eta = 2 \times 10^{-3}, 5 \times 10^{-4}, 1.25 \times 10^{-4} ,
  3.1 \times 10^{-5}$, and $7.7 \times 10^{-6}$, from left to right curves respectively. Fitted exponential laws are also added over the last three curves
  scaling as $\sim e^{0.45t/t_A}$.
  }
\label{figu7}
\end{figure}

 \begin{figure}
\centering
 \includegraphics[scale=0.7]{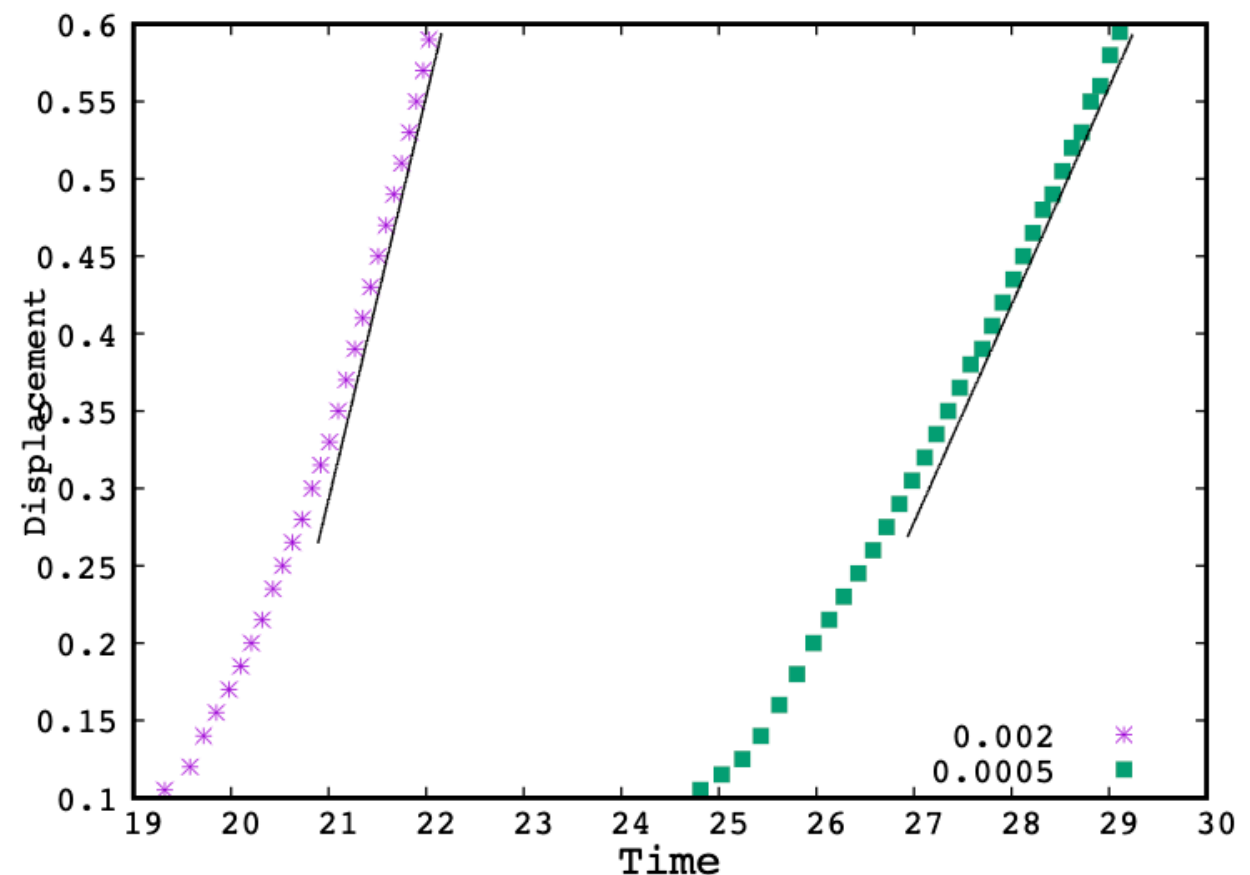}
  \caption{Displacement of the magnetic axis of the plasma core $\xi_0 (t)/a$ versus time (normalized with respect to $t_A$) for two values of the resistivity, $\eta = 2 \times 10^{-3}$ and
  $5 \times 10^{-4}$. The magnetic Prandtl number is $P_r= 1$. For relatively large values of the displacement, a linear fit can be done in order to estimate the speed of the reconnection.
  }
\label{figu8}
\end{figure}

 \begin{figure}
\centering
 \includegraphics[scale=0.7]{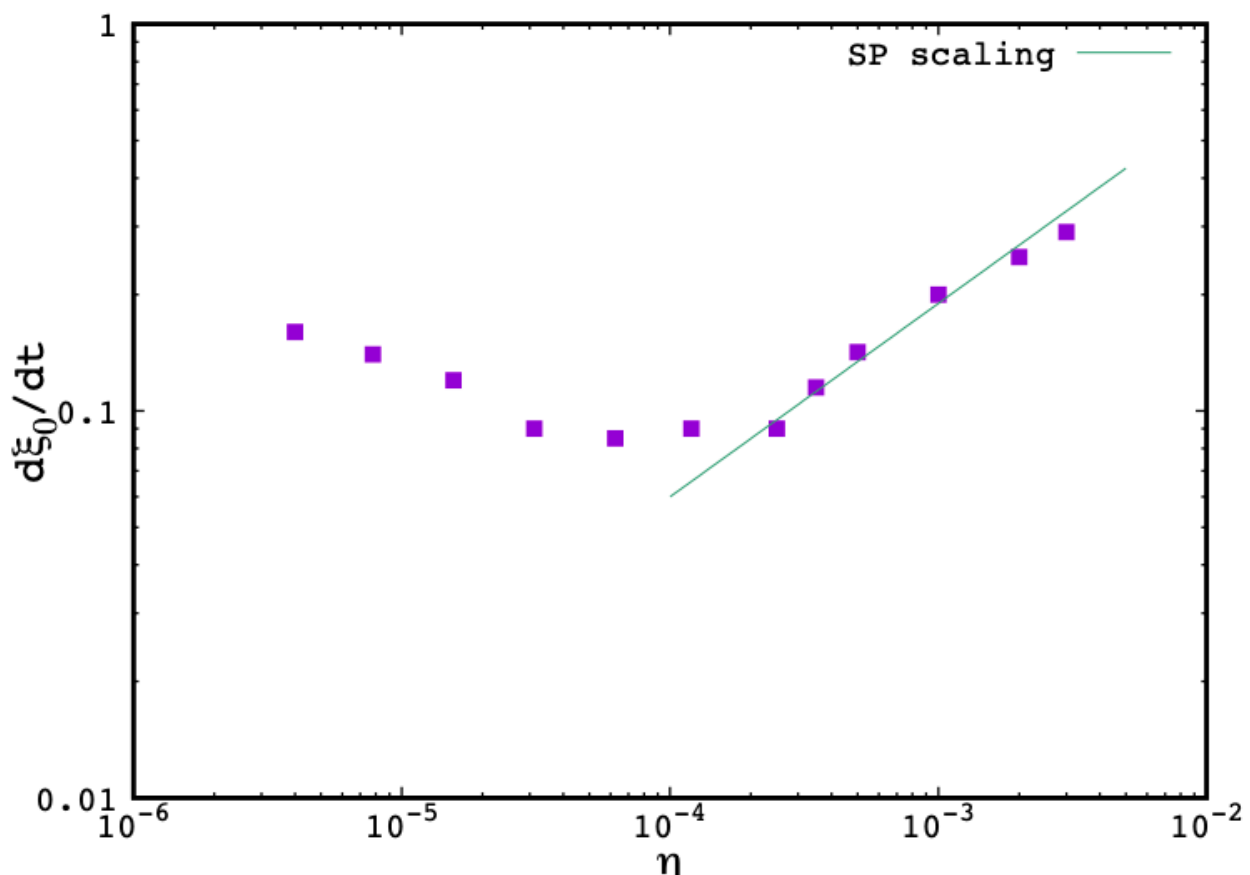}
  \caption{Time derivative of the magnetic axis displacement $\xi_0 (t)$ (see Figure 8) versus the resistivity value. The Sweet-Parker (SP)
  scaling as $\eta^{1/2}$ is also plotted for comparison.
  }
\label{figu9}
\end{figure}

  \begin{figure}
\centering
 \includegraphics[scale=0.18]{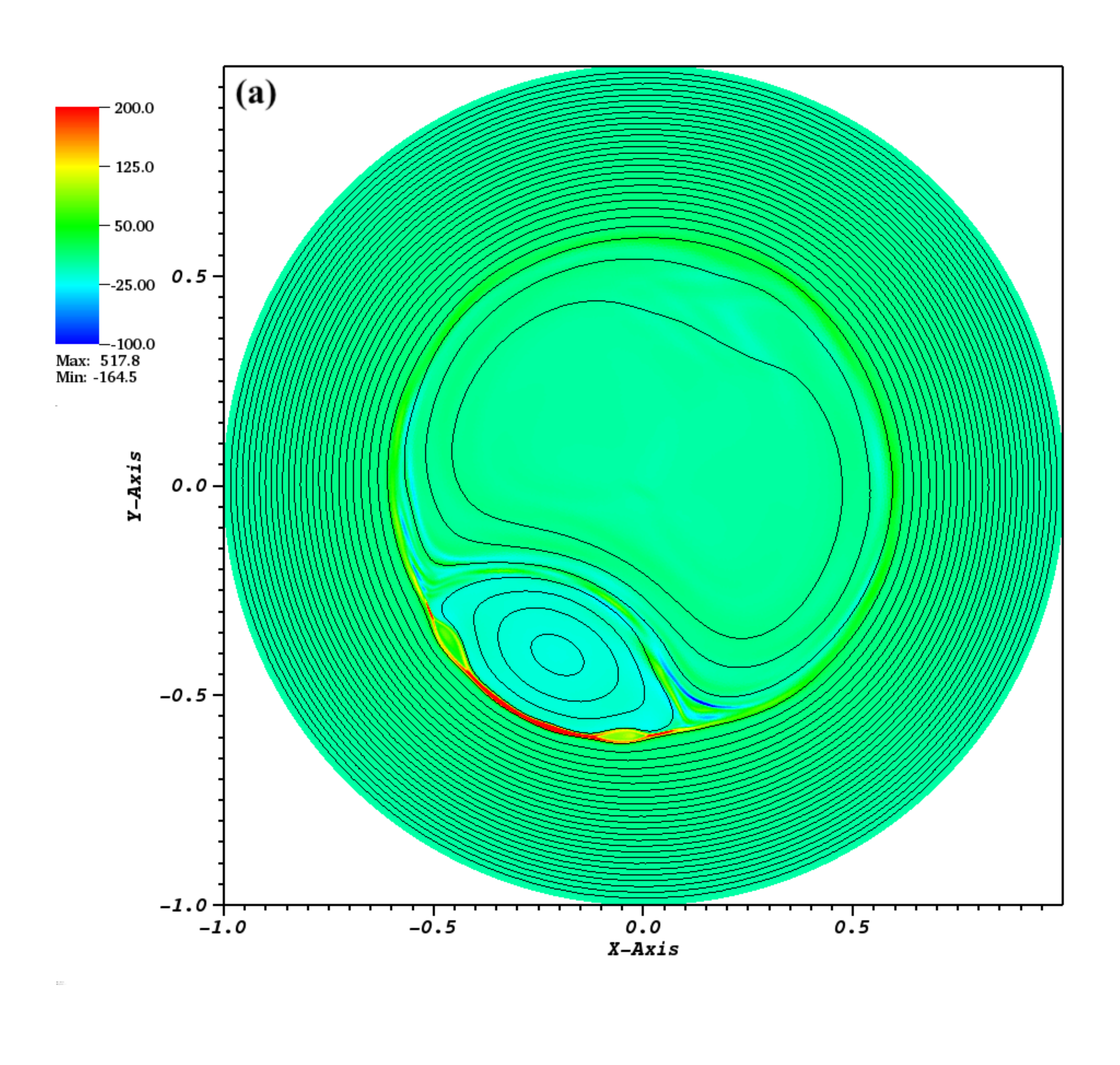}
 \includegraphics[scale=0.18]{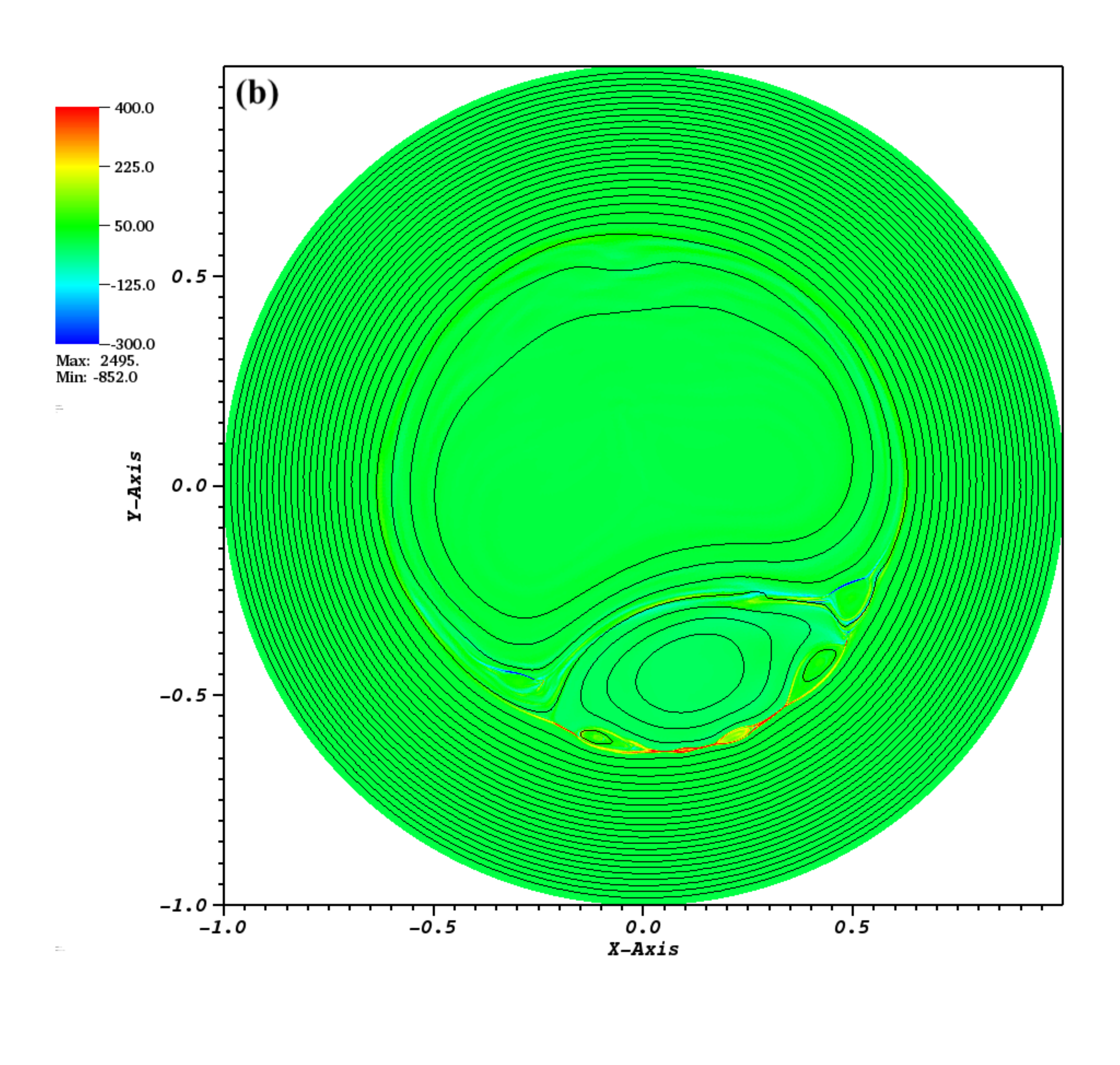}
 \includegraphics[scale=0.18]{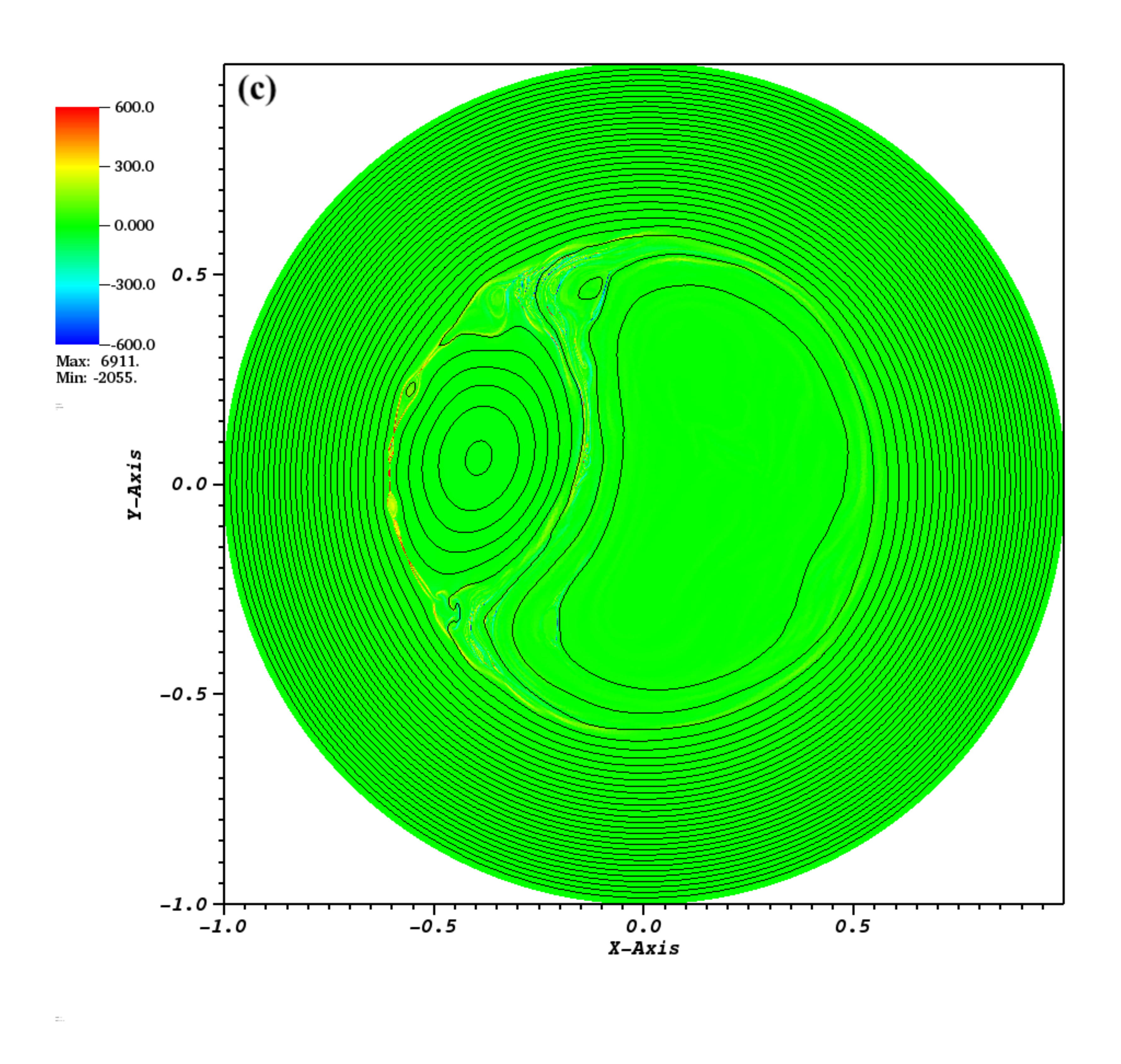}
 \includegraphics[scale=0.27]{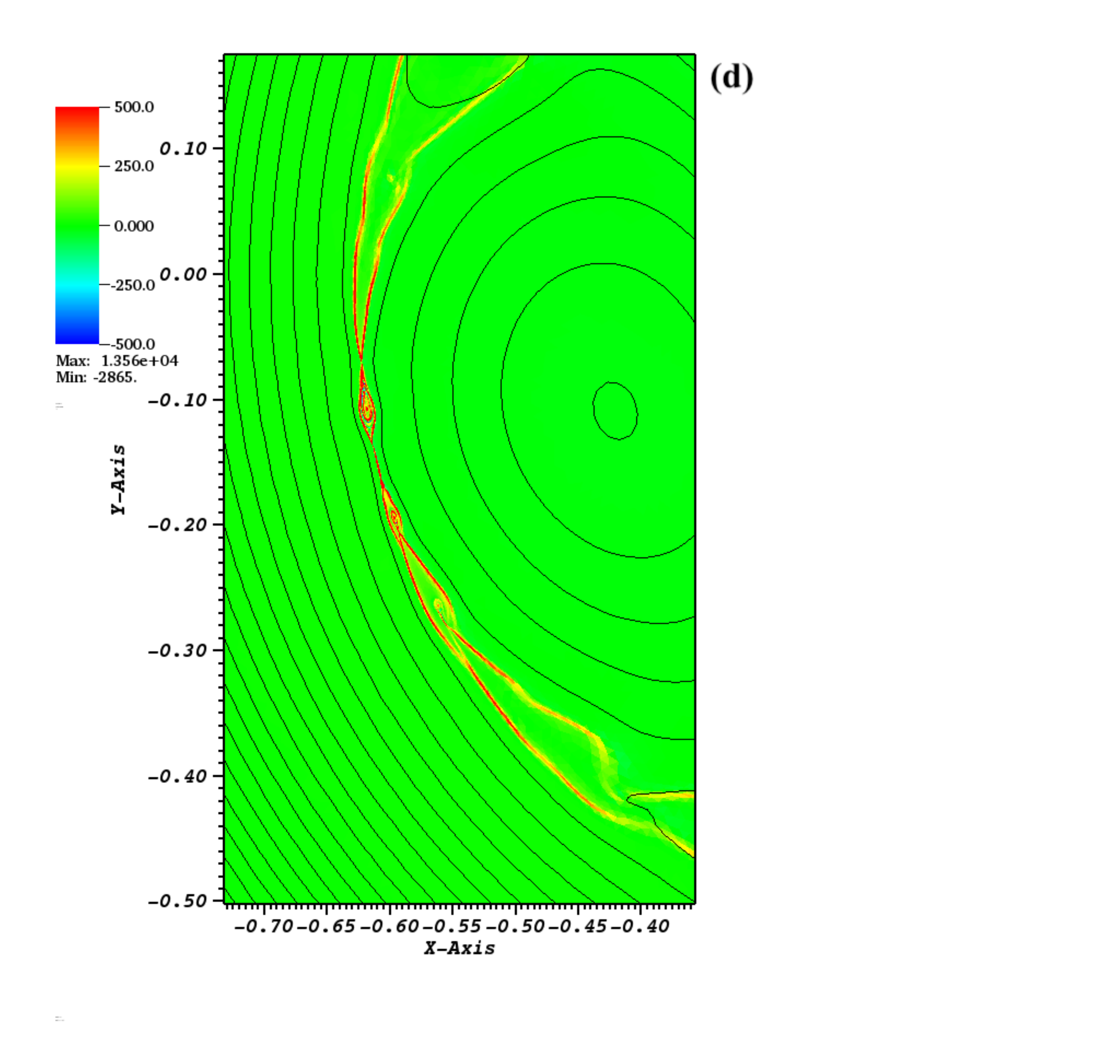}
  \caption{Colored contour maps of the current density overlaid with magnetic field lines for simulations using $\eta = 1.2 \times 10^{-4}$ (a-panel),
  $\eta = 3.12 \times 10^{-5}$ (b-panel), $\eta = 7.7 \times 10^{-6}$ (c-panel), and $\eta = 3.9 \times 10^{-6}$ (zoom on the current sheet in d-panel).
  The four panels correspond to the same current reconnection state of $\xi_0 (t)  \simeq 0.4a$.
   Saturated current density values are used to facilitate the visualization.
   }
\label{figu10}
\end{figure}

 \begin{figure}
\centering
 \includegraphics[scale=0.6]{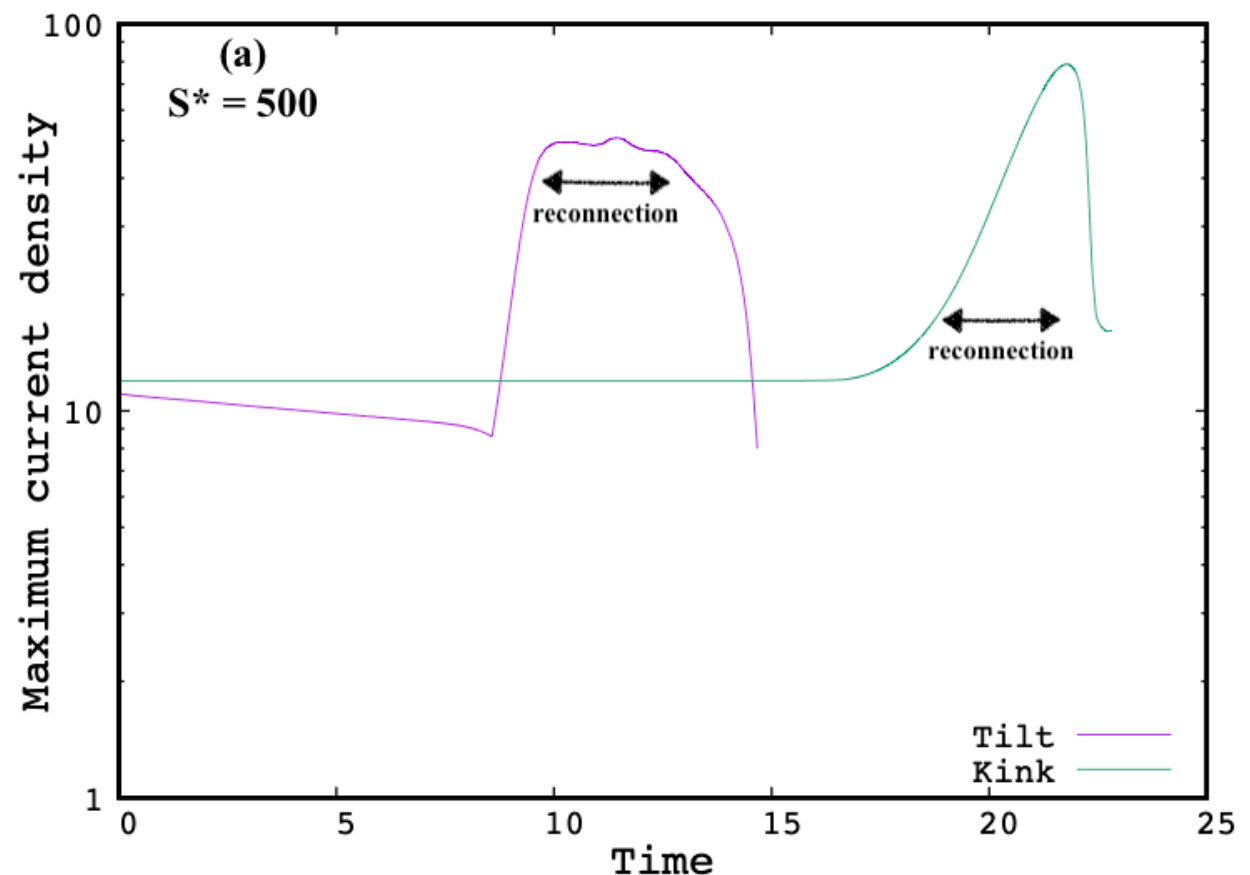}
 \includegraphics[scale=0.6]{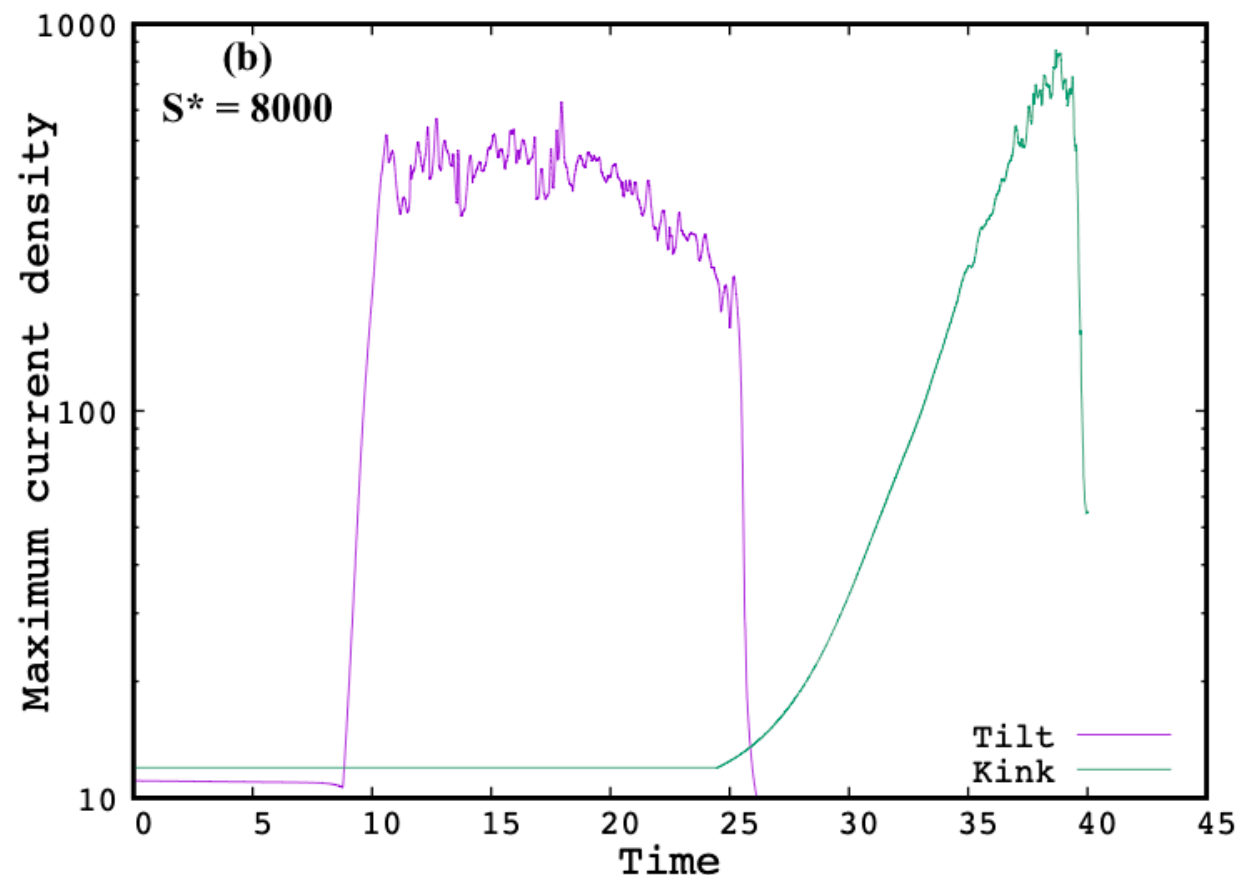}
 \includegraphics[scale=0.6]{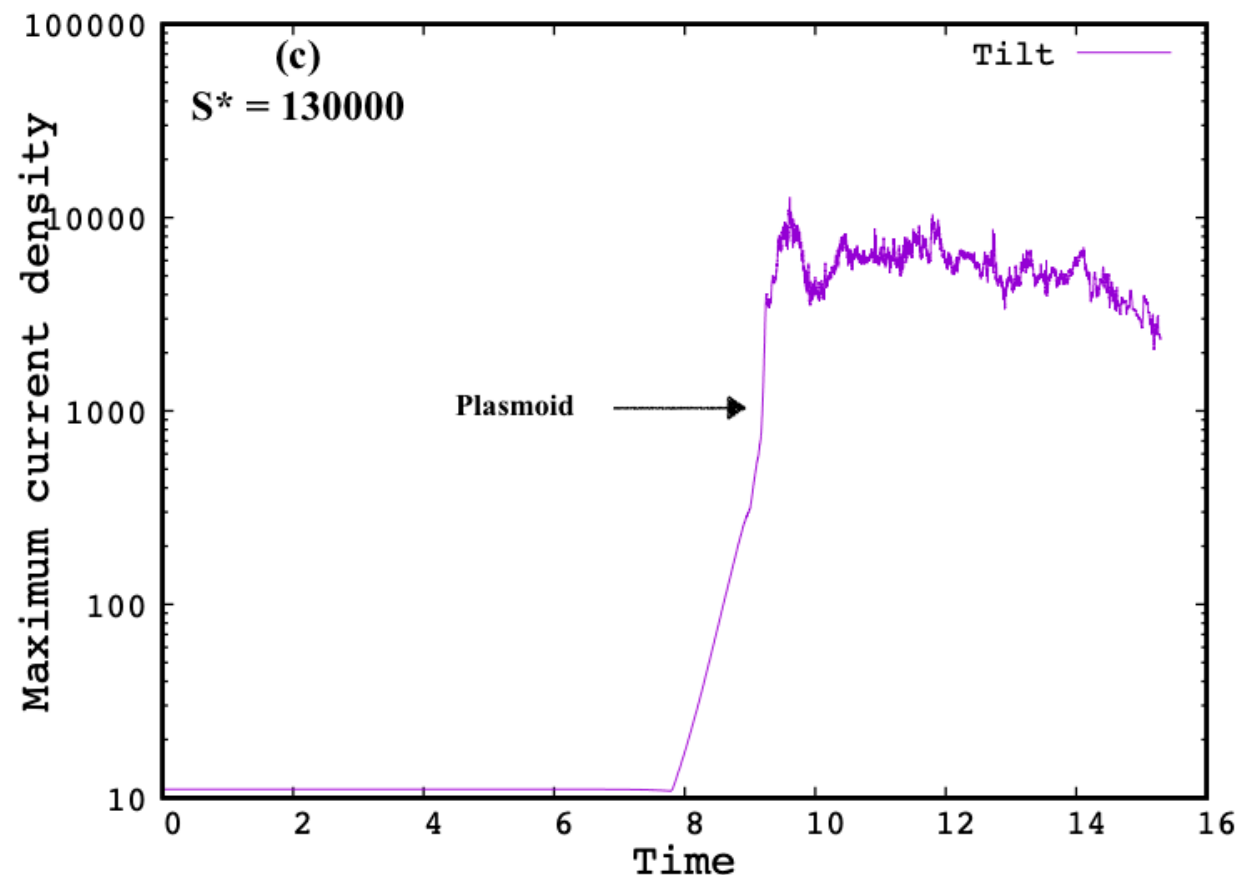}
 \includegraphics[scale=0.6]{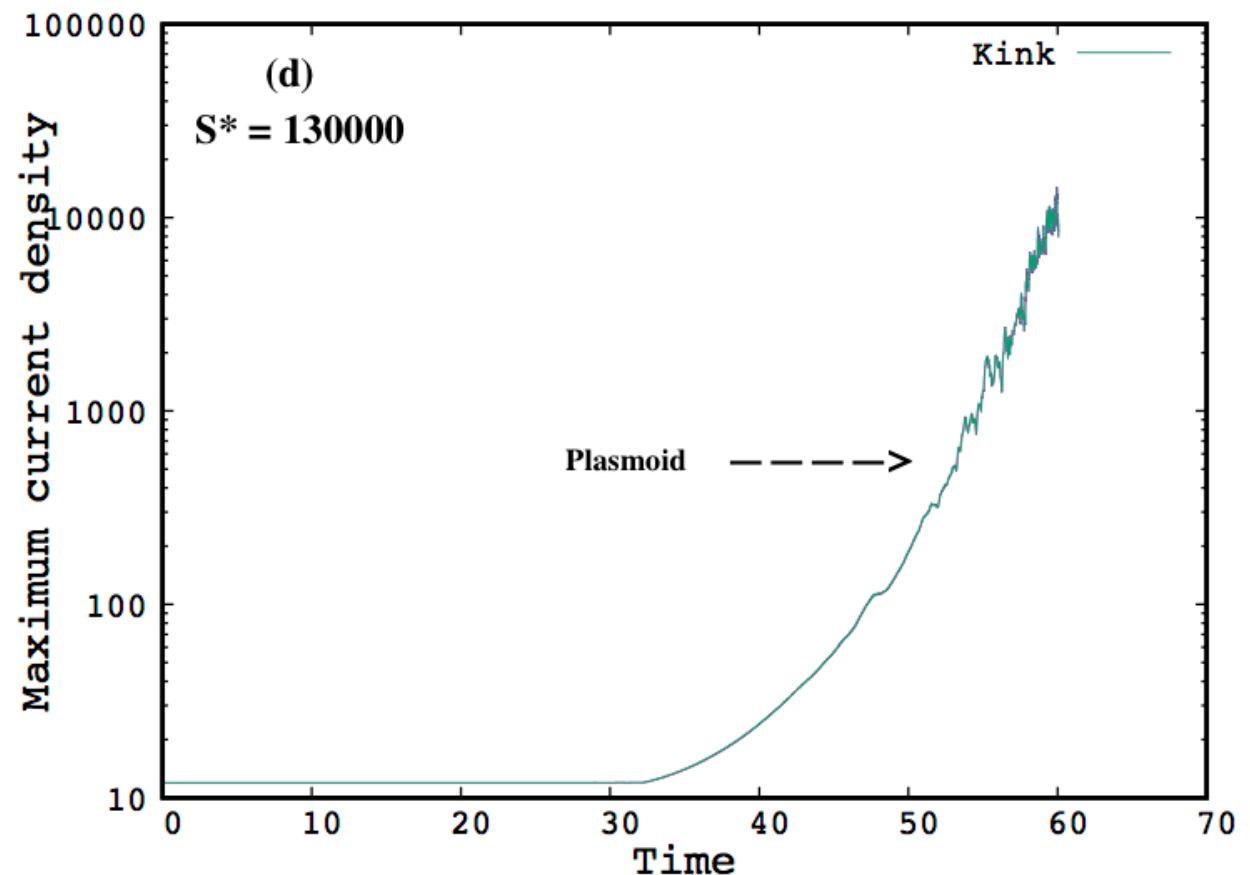}
  \caption{Maximum current density as a function of time for tilt and resistive kink instabilities. (a) Evolution for tilt/kink obtained for an inverse resistivity
  value $S^* = 500$ (or equivalently $\eta = 2 \times 10^{-3}$). (b) Evolution for tilt/kink for $S^* = 8000$ (or equivalently $\eta = 1.25 \times 10^{-4}$).
  (c) Evolution for tilt for $S^* = 1.3  \times 10^5$  (or equivalently  $\eta = 7.7 \times 10^{-6}$). (d) Evolution for kink for $S^* = 1.3  \times 10^5$.
   The early formation of plasmoids is indicated using
  the arrow.
   }
\label{figu11}
\end{figure}

 \begin{figure}
\centering
 \includegraphics[scale=0.27]{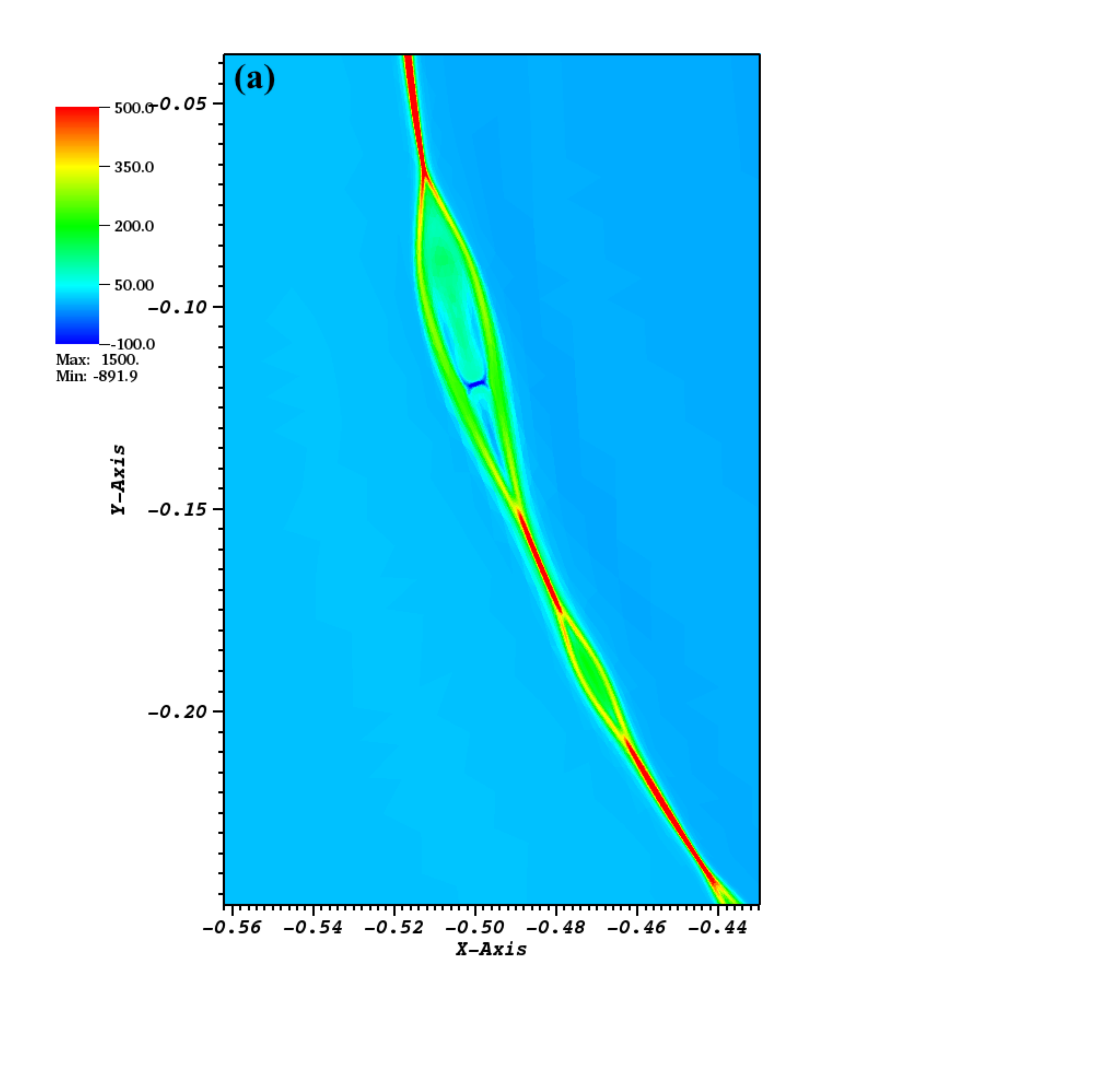}
 \includegraphics[scale=0.27]{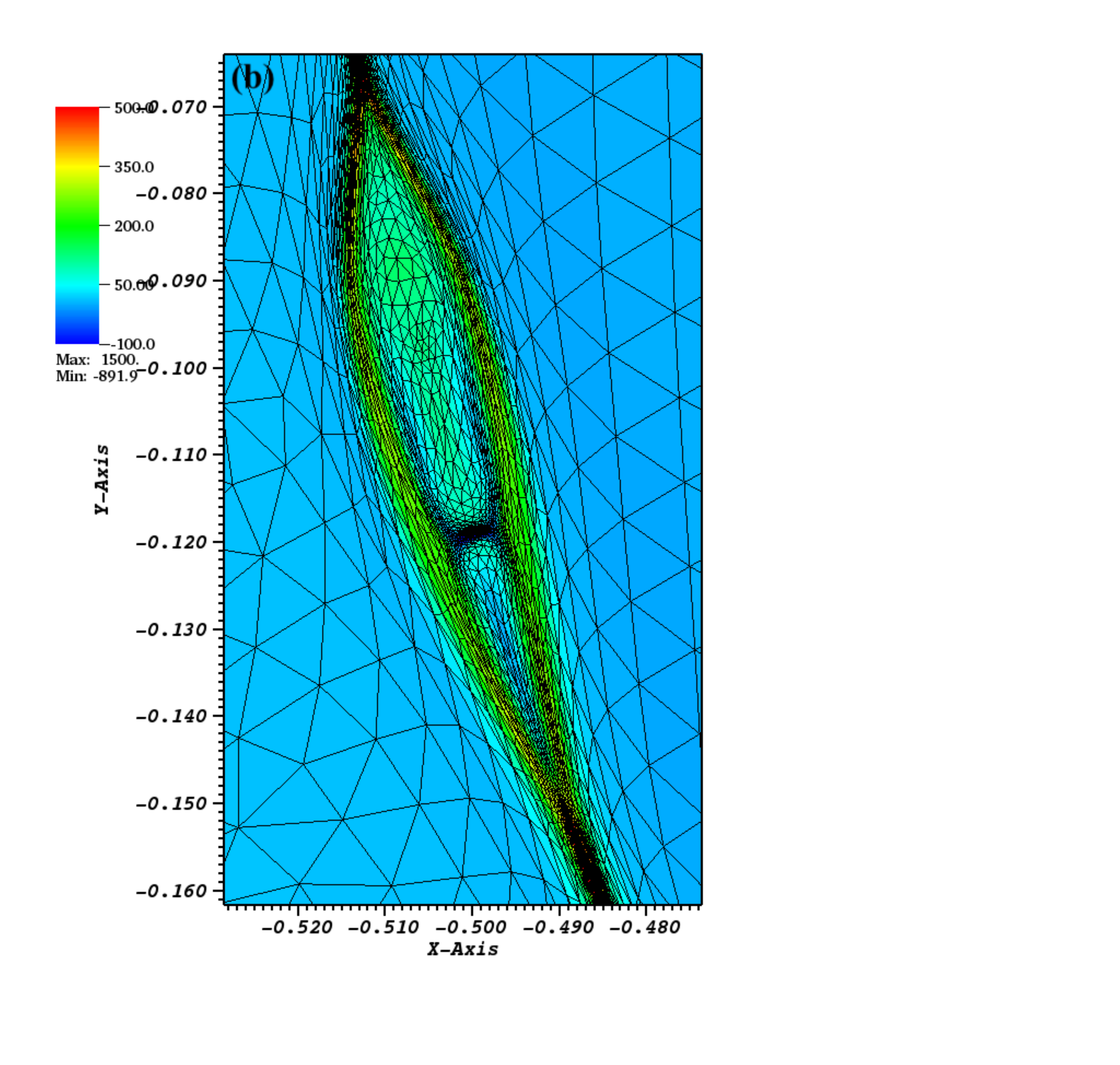}
   \caption{(a) Zoom-in on a portion of the current layer invaded by plasmoids during reconnection, using a saturated colored contour map
   of the current density. (b) Further zoom-in on panel (a) showing the coalescence event between the two upper plasmoids overlaid with the grid.
      }
\label{figu12}
\end{figure}

 \begin{figure}
\centering
 \includegraphics[scale=0.63]{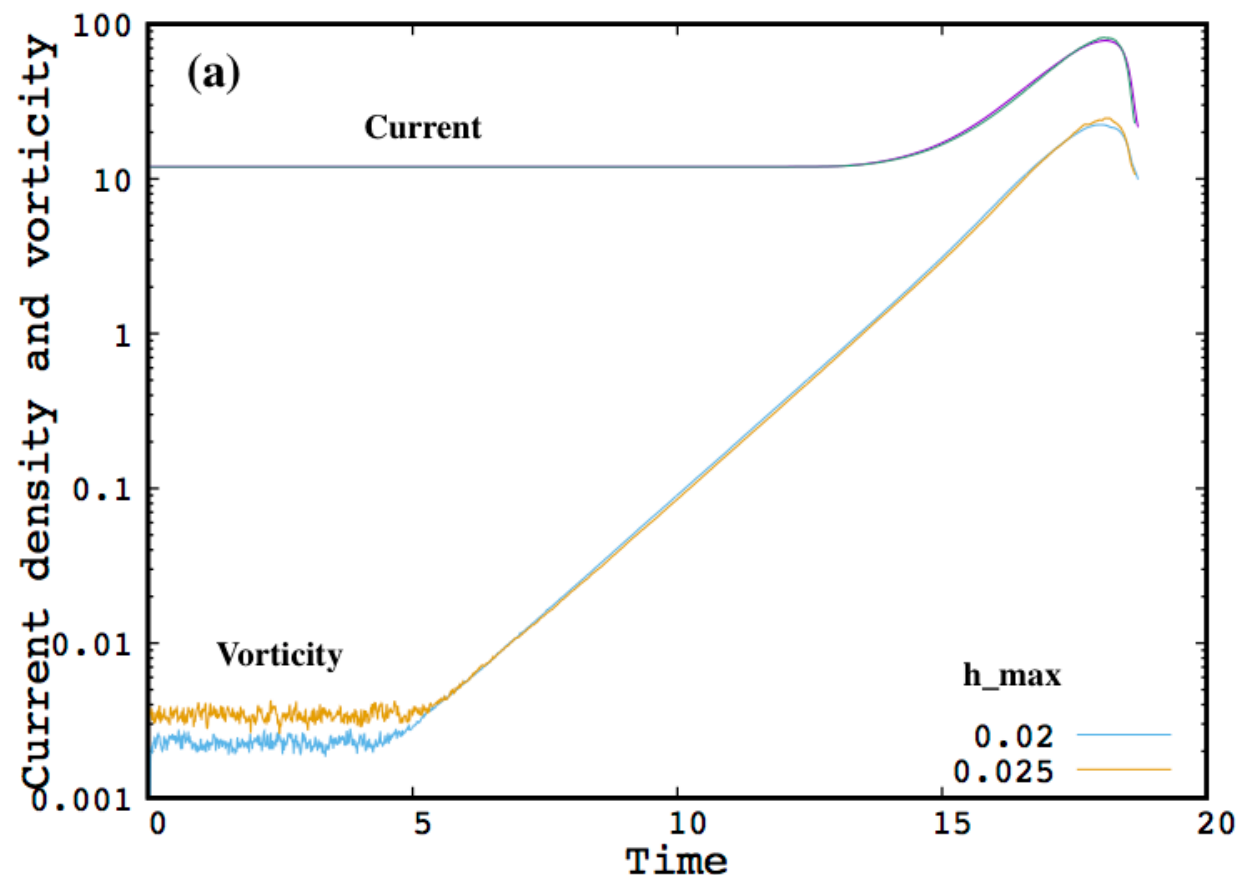}
 \includegraphics[scale=0.63]{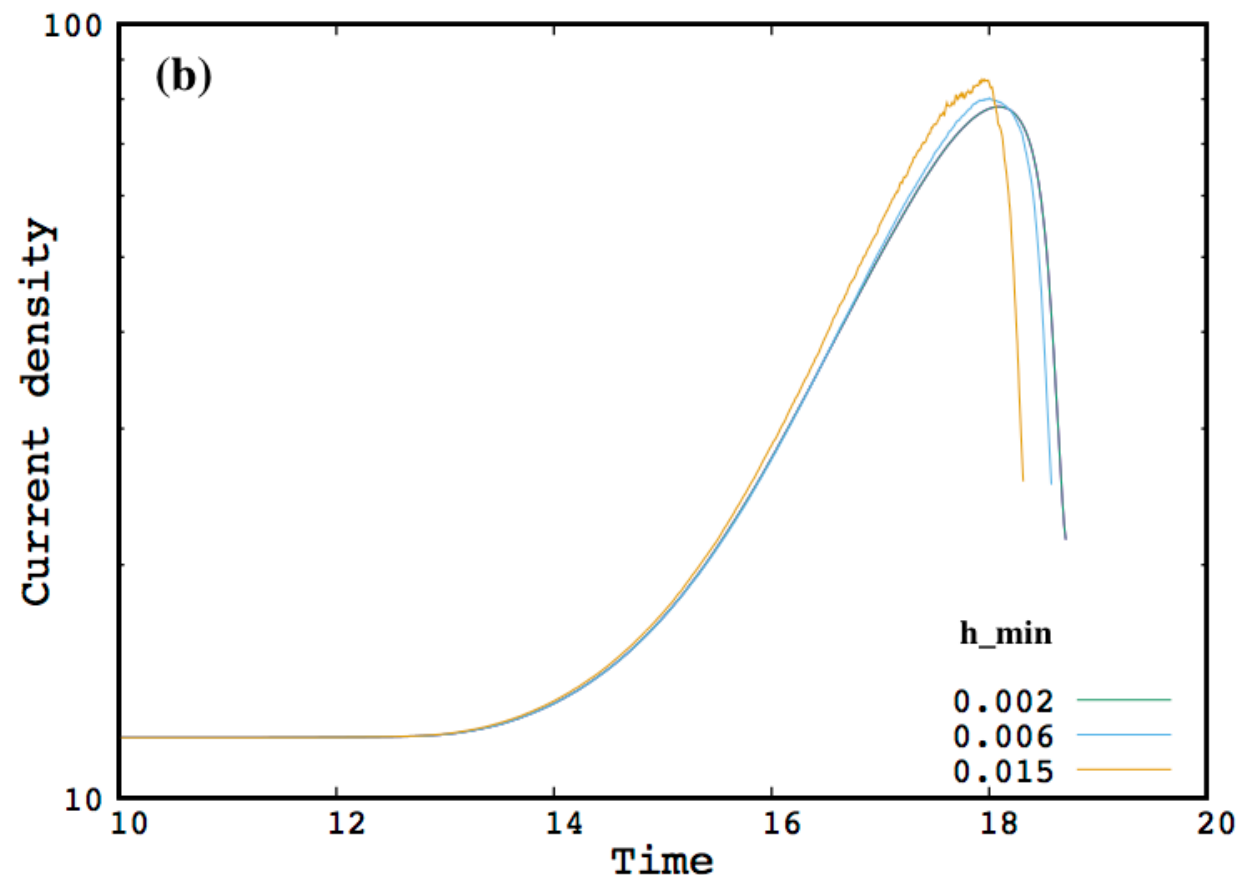}
   \caption{(a) Mesh convergence with the maximum edge size $h_{max}$ for the results (i.e. maximum current density and maximum vorticity) plotted in Fig. 3
   for SP regime, where two different values of $h_{max}$ are employed (e.g. $0.02$ and $0.025$). (b) Mesh convergence with the minimum edge size $h_{min}$
   for the same resistivity case using $h_{max} = 0.02$,
    where three different values of $h_{min}$ are employed, i.e. $2 \times 10^{-3}$, $6 \times 10^{-3}$, and $1.5 \times 10^{-2}$.
      }
\label{figu13}
\end{figure}

 \begin{figure}
\centering
 \includegraphics[scale=0.63]{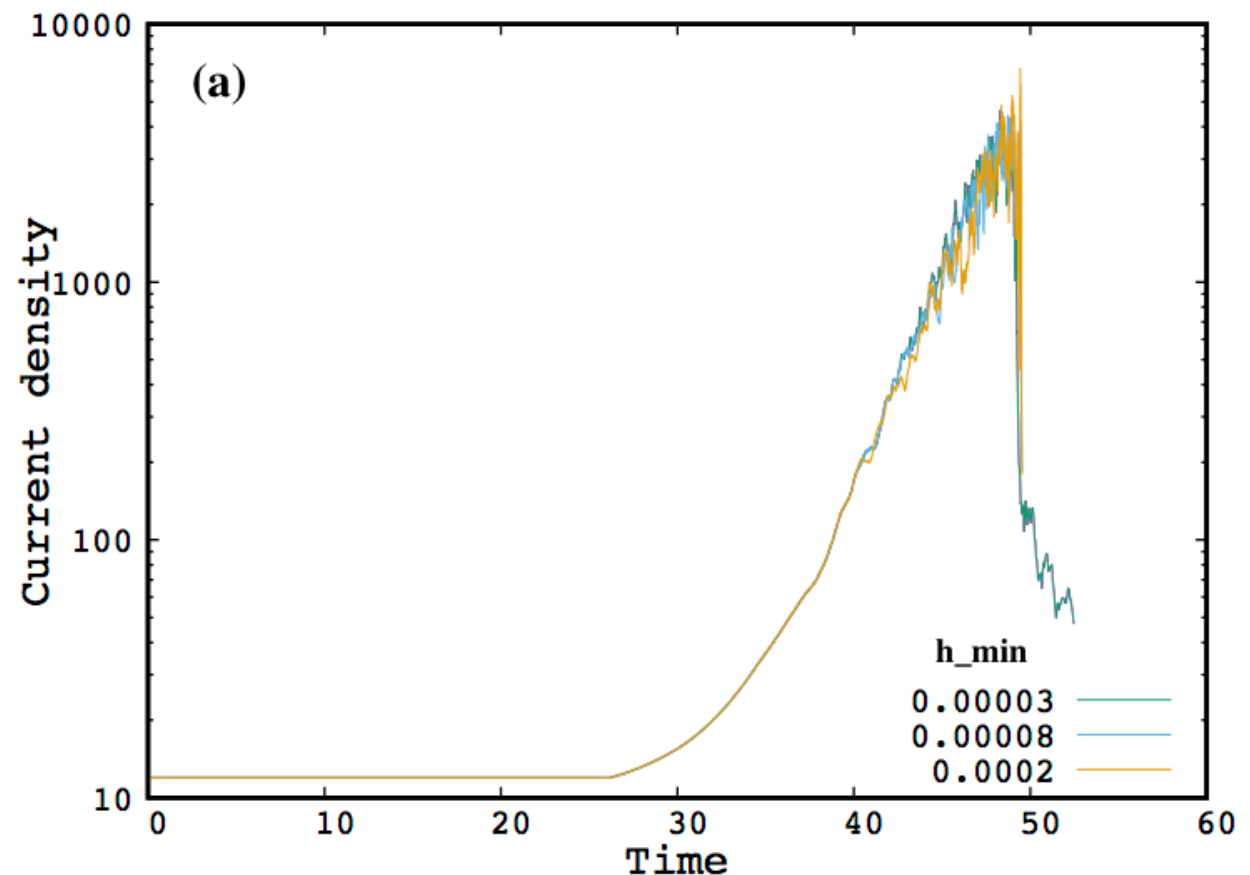}
 \includegraphics[scale=0.63]{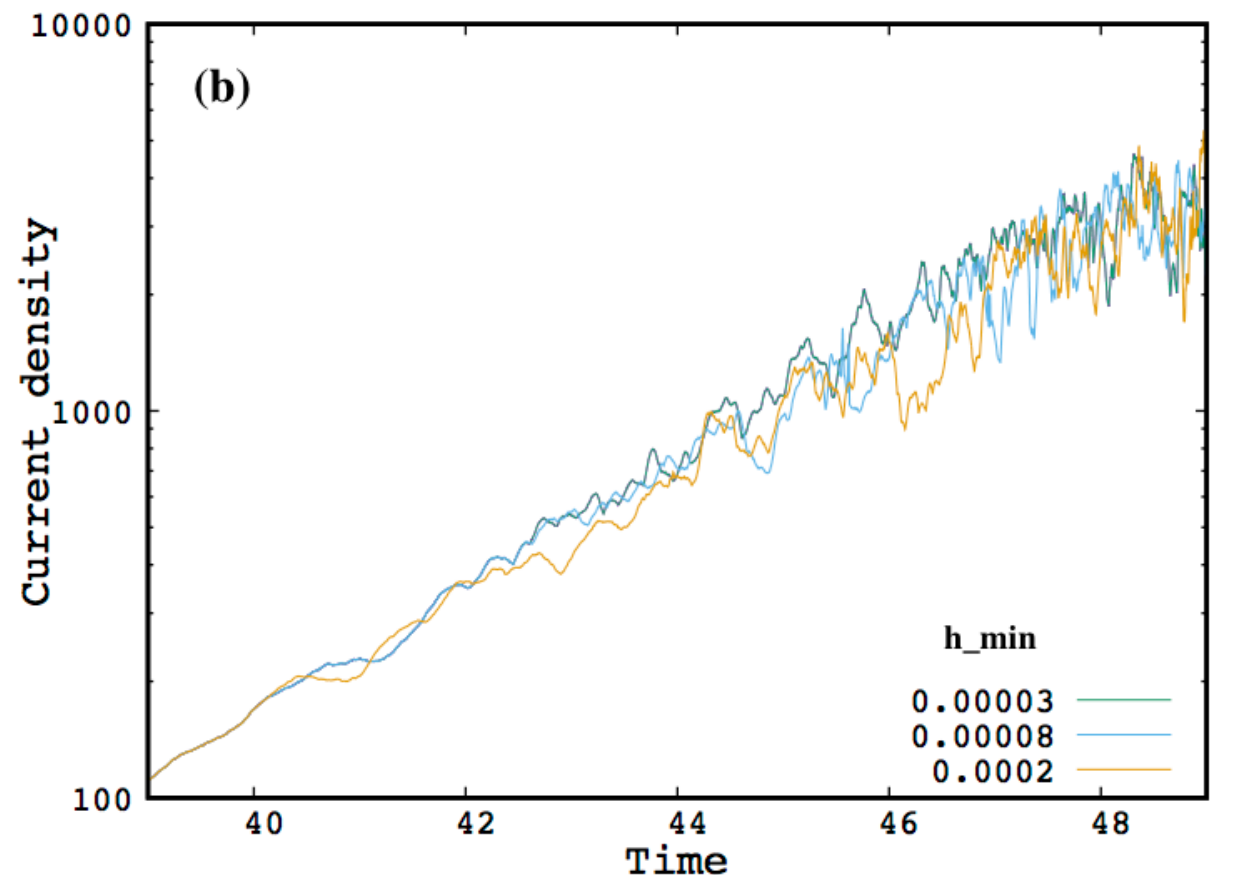}
   \caption{(a) Mesh convergence with the minimum edge size $h_{min}$ for the maximum current density evolution plotted in Fig. 7 for $\eta = 3.1 \times 10^{-5}$ and
   employing $h_{max} = 0.01$.
   Three different values of $h_{min}$ are employed, i.e. $3 \times 10^{-5}$, $8 \times 10^{-5}$, and $2 \times 10^{-4}$. (b) Zoom-in on panel (a) for a time interval
  mainly restricted to the plasmoid phase.
         }
\label{figu14}
\end{figure}


\begin{thebibliography}{10}

\bibitem{priest00}
E.R.~{Priest} and T.G.~{Forbes},
\newblock {\em {Magnetic Reconnection}}
\newblock (Cambridge University Press, 2000).
https://doi.org/10.1017/CBO9780511525087

\bibitem{bisk09}
D.~{Biskamp},
\newblock {\em {Nonlinear Magnetohydrodynamics}}
\newblock (Cambridge University Press, 2009).
https://doi.org/10.1017/CBO9780511599965

\bibitem{hua10}
Y.M.~{Huang}, and A.~{Bhattacharjee},
\newblock {Phys. Plasmas} \textbf{17}, 062104 (2010).
https://doi.org/10.1063/1.3420208

\bibitem{richard90} 
R.L.~{Richard}, R.D.~{Sydora}, and M.~{Ashour-Abdalla},
\newblock {Phys. Fluids B} \textbf{2}, 488 (1990).
https://doi.org/10.1063/1.859338

\bibitem{lou07} 
N.F.~{Loureiro}, A.A.~{Schekochihin}, and S.~C.~{Cowley},
\newblock {Phys. Plasmas} \textbf{14}, 100703, (2007).
https://doi.org/10.1063/1.2783986

\bibitem{par57} 
E.N.~{Parker},
\newblock {Journal of Geophysical Research} \textbf{62}, 50520 (1957).
https://doi.org/10.1029/JZ062i004p00509

\bibitem{kado75} 
B.B.~{Kadomstev},
\newblock {Fizika Plazmy} \textbf{1}, 710 (1975).

\bibitem{hua13}
Y.M.~{Huang}, and A.~{Bhattacharjee},
\newblock {Phys. Plasmas} \textbf{20}, 055702 (2013).
https://doi.org/10.1063/1.4802941


\bibitem{ni12}
L.~Ni, U.~{Ziegler}, Y.M.~{Huang}, J.~{Lin}, and Z.~{Mei},
\newblock {Phys. Plasmas} \textbf{19},072902, (2012).
https://doi.org/10.1063/1.4736993

\bibitem{comi16}
L.~{Comisso}, and D.~{Grasso},
\newblock {Phys. Plasmas} \textbf{23},032111, (2016).
https://doi.org/10.1063/1.4942940

\bibitem{hua17}
Y.M.~{Huang}, L.~{Comisso}, and A.~{Bhattacharjee},
\newblock {The Astrophysical Journal} \textbf{849}, 75 (2017).
https://doi.org/10.3847/1538-4357/aa906d


\bibitem{baty19} 
H.~{Baty},
\newblock {The Astrophysical Journal Supplement Series} \textbf{243}, 23 (2019).
https://doi.org/10.3847/1538-4365/ab2cd2

\bibitem{bat20a} 
H.~{Baty},
Formation of plasmoid chains and fast magnetic reconnection during nonlinear evolution of the tilt instability. (2020)
https://arxiv.org/abs/2001.07036

\bibitem{bat20b} 
H.~{Baty},
On the growth rate of plasmoid chains during nonlinear viscoresistive evolution of the tilt instability. (2020)
https://arxiv.org/abs/2003.08660

\bibitem{rip19}
B.~{Ripperda}, O.~{Porth}, L.~{Sironi}, and R.~{Keppens},
\newblock {MNRAS}    \textbf{485}, 299, (2019).
https://doi.org/10.1093/mnras/stz387


\bibitem{com16}
L.~{Comisso}, M.~{Lingam}, M.~{Huang}, and A.~{Bhattacharjee},
\newblock {Phys. Plasmas} \textbf{23}, 100702 (2016).
https://doi.org/10.1063/1.4964481

\bibitem{com17}
L.~{Comisso}, M.~{Lingam}, M.~{Huang}, and A.~{Bhattacharjee},
\newblock {The Astrophysical Journal} \textbf{850}, 142 (2017).
https://doi.org/10.3847/1538-4357/aa9789  

\bibitem{puc14}
F.~{Pucci} and M.~{Velli},
\newblock {Astrophys. J.} \textbf{780}, L19 (2014).
\newline
https://doi.org/10.1088/2041-8205/780/2/L19

\bibitem{puc18}
F.~{Pucci}, M.~{Velli},and A.~{Tenerani},
\newblock {Phys. Plasmas} \textbf{25}, 032113  (2018).
https://doi.org/10.1063/1.5022988

\bibitem{landi15} 
S.~{Landi}, L.~{Del Zanna}, E.~{Papini}, F.~{Pucci}, and M.~{Velli},
\newblock {The Astrophysical Journal} \textbf{806}, 131 (2015).
https://doi.org/10.1088/0004-637X/806/1/131

\bibitem{white91} 
R.B.~{White},
\newblock {Rev. Mod. Phys.} \textbf{58}, 183 (1991).
https://doi.org/10.1103/RevModPhys.58.183

\bibitem{uzd10} 
D.A.~{Uzdensky}, N.F.~{Loureiro}, and A.A.~{Schekochihin},
\newblock {Phys. Rev. Lett.} \textbf{105}, 235002 (2010).
https://doi.org/10.1103/PhysRevLett.105.235002

\bibitem{lou12} 
N.F.~{Loureiro}, N.~F., Samtaney, and D.A.~{Uzdensky},
\newblock {Phys. Plasmas} \textbf{19}, 042303 (2012).
https://doi.org/10.1063/1.3703318

\bibitem{ji11}
H.~{Ji}, and W.~{Doughton},
\newblock {Phys. Plasmas} \textbf{18}, 11207 (2011).
https://doi.org/10.1063/1.3647505

\bibitem{shi01} 
K.~{Shibata}, and S.~{Tanuma},
\newblock {Earth, Planets and Space} \textbf{53}, 473 (2001).
https://doi.org/10.1186/BF03353258

\bibitem{bat20c} 
H.~{Baty},
Petschek-type reconnection in the high-Lundquist-number regime during nonlinear evolution on the tilt instability. (2020)
https://arxiv.org/abs/2005.04221

\bibitem{huy01} 
G.T.A.~{Huysmans}, T.C.~{Hender}, N.C.~{Hawkes}, and X.~{Litaudon},
\newblock {Phys. Rev. Lett.} \textbf{87}, 245002 (2001).
https://doi.org/10.1103/PhysRevLett.87.245002

\bibitem{cza08} 
 O.~{Czarny}, and G.T.A.~{Huysmans},
\newblock {Journal of Computational Physics} \textbf{227}, 7423 (2008).
https://doi.org/10.1016/j.jcp.2008.04.001

\bibitem{hec12} 
F.~{Hecht},
\newblock {Journal of Numerical Mathematics} \textbf{20}, 251 (2012).
https://doi.org/10.1515/jnum-2012-0013

\bibitem{men18} 
J.~{Mendonca}, D.~{Chandra}, A.~{Sen}, and A.~{Thyagaraja}, 
\newblock {Phys. Plasmas} \textbf{25}, 022504 (2018).
https://doi.org/10.1063/1.5009506

\bibitem{bat91} 
H.~{Baty}, J.F.~{Luciani}, and M.N.~{Bussac},
\newblock {Nuclear Fusion} \textbf{31}, 2055 (1991).
https://doi.org/10.1088/0029-5515/31/11/002

\bibitem{mei12}
Z.~{Mei}, C.~{Shen}, N.~{Wu}, J.~{Lin}, N.A.~{Murphy}, and I.I.~Roussev,
\newblock {MNRAS} \textbf{425}, 2824 (2012).
https://doi.org/10.1111/j.1365-2966.2012.21625.x

\bibitem{ni16}
L.~{Ni}, L.~{Lin}, N.~{Wu}, I.I.~Roussev, and B.~Schmieder,
\newblock {The Astrophysical Journal} \textbf{832}, 195 (2016).
https://doi.org/10.3847/0004-637X/832/2/195

\bibitem{hua19} 
Y.M.~{Huang}, L.~{Comisso}, and A.~{Bhattacharjee},
\newblock {Phys. Plasmas} \textbf{26}, 092112 (2019).
https://doi.org/10.1063/1.5110332

\bibitem{tol18} 
E.A.~{Tolman}, N.F.~{Loureiro}, and D.A.~{Uzdensky},
\newblock {Journal of Plasma Physics} \textbf{84}, 905840115 (2018).
https://doi.org/10.1017/S002237781800017X

\bibitem{com20} 
L.~{Comisso} (private communication, 2020)

\bibitem{dong18}
C.~{Dong}, L.~{Wang}, Y.M.~{Huang}, L.~{Comisso}, and A.~{Bhattacharjee},
\newblock {Phys. Rev. Lett.}  \textbf{121}, 165101 (2018).
https://doi.org/10.1103/PhysRevLett.121.165101

\bibitem{stri16} 
E.~{Striani}, A.~{Mignone}, B.~{Vaidya}, G.~{Bodo}, and A.~{Ferrari}, 
\newblock {MNRAS} \textbf{462}, 2970 (2016).
https://doi.org/10.1093/mnras/stw1848

\bibitem{gun15} 
S.~{G\"unter}, Q.~{Yu}, K.~{Lackner}, A.~{Bhattacharjee}, and Y.M.~{Huang},
\newblock {Plasma Phys. Control. Fusion} \textbf{57}, 014017 (2015).
https://doi.org/10.1088/0741-3335/57/1/014017

\bibitem{hua11} 
Y.M.~{Huang}, A.~{Bhattacharjee}, and B.~P. A. ~{Sullivan},
\newblock {Phys. Plasmas} \textbf{18}, 072109 (2011).
https://doi.org/10.1063/1.3606363



\end{thebibliography}
\end{document}